\documentclass[runningheads]{llncs}
\usepackage[T1]{fontenc}
\usepackage{newtxtext}       %
\usepackage[varvw]{newtxmath}       

\usepackage{algorithm}
\usepackage{hyperref}
\usepackage{mathpartir}
\usepackage{caption}
\usepackage{times}
\usepackage[noend]{algpseudocode}
\usepackage{subcaption}
\usepackage{adjustbox}
\usepackage{booktabs}
\usepackage{makecell}
\usepackage{float}
\usepackage{orcidlink}
\usepackage{xspace}

\newcommand\SPC{ScenicProver\xspace}

\newcommand{\customparagraph}[1]{
    \vspace{1ex}%
    \noindent%
    \emph{#1}%
}

\newtheorem{assum}{Assumption}

\usepackage{graphicx}
\begin{document}
\title{\SPC{}: A Framework for Compositional Probabilistic Verification of Learning-Enabled Systems}

%
\titlerunning{\SPC{}}
%
\author{
Eric Vin\inst{1}\orcidlink{0000-0002-3089-1129} \and
Lucy Horowitz\inst{3}\orcidlink{0009-0008-4480-6934}\and
Kyle A. Miller\inst{1}\orcidlink{0000-0001-7400-5304} \and\\
Inigo Incer\inst{2} \orcidlink{0000-0001-7933-692X} \and
Sanjit A. Seshia\inst{3}\orcidlink{0000-0001-6190-8707} \and 
Daniel J. Fremont\inst{1}\orcidlink{0000-0002-9992-9965}
}
\institute{
University of California, Santa Cruz, Santa Cruz CA 95064, USA\\
\email{\{evin, kymiller, dfremont\}@ucsc.edu}  \and
University of Michigan, Ann Arbor MI 48109, USA\\
\email{iir@umich.edu} \and
University of California, Berkeley, Berkeley CA 94720, USA\\
\email{\{lucyhorowitz, sseshia\}@berkeley.edu}
}
\authorrunning{E. Vin et al.}
\maketitle              
\begin{abstract}
Full verification of learning-enabled cyber-physical systems (CPS) has long been intractable due to challenges including black-box components and complex real-world environments. Existing tools either provide formal guarantees for limited types of systems or test the system as a monolith, but no general framework exists for compositional analysis of learning-enabled CPS using varied verification techniques over complex real-world environments.
This paper introduces a tool, ScenicProver, that aims to fill this gap by enabling the construction of sound, formally-checked assurance cases for CPS. ScenicProver is backed by a formal theory, which we lay out in this paper. ScenicProver supports: (1) environment modeling using the Scenic probabilistic programming language; (2) compositional system modeling with clear component interfaces, ranging from interpretable code to black boxes; (3) assume-guarantee contracts over those components using an extension of Linear Temporal Logic containing arbitrary Scenic expressions; (4) evidence generation through testing, formal proofs in Lean 4, and external assumptions; (5) sound combination of generated evidence using contract operators with Pacti; and (6) automatic generation of assurance cases tracking the provenance of system-level guarantees.
We demonstrate its effectiveness through two case studies: one on an automatic emergency braking system with sensor fusion and one on drone safety. By leveraging manufacturer guarantees for sensors and focusing testing effort on uncertain conditions, our approach enables stronger probabilistic guarantees than monolithic testing with the same computational budget.

\end{abstract}
\section{Introduction}

Full verification of learning-enabled cyber-physical systems (LE-CPS) has long been a goal of the formal methods community, but has proven intractable for a variety of reasons. First and foremost is the difficulty in reasoning about black-box or unstructured components such as machine learning (ML) models~\cite{TowardsVerifiedAI_Seshia_22}. It is also difficult to model diverse real-world environments in a format that is conducive to existing verification techniques. Finally, small changes to the system, such as those occurring when an ML component is trained or fine-tuned, can invalidate all analysis done so far.
However, these problems primarily arise when treating the system as a \emph{monolith}, with the goal of a proof that a specification is satisfied under \emph{all possible environments}. If one is willing to relax these assumptions, the problem becomes more feasible.

One route is to settle for a weaker guarantee: rather than prove correctness in all possible environments, we can systematically test the system via falsification~\cite{VerifAI_Dreossi_19,AdaptiveStressTesting_Koren_18}, obtain a probabilistic guarantee from simulations as in statistical model checking~\cite{StatisticalModelChecking_Legay_10}, or even make a reasonable assumption based on manufacturer specifications or real-world testing~\cite{AssuranceCases_Rushby_2015}. While such evidence of correctness is not as ironclad as a formal proof, it can still be useful: in fact, such reasoning is commonplace in \emph{assurance cases} used to justify safety in certification processes for civil aviation, defense, and infrastructure (e.g., FAA DO-178C~\cite{AssuranceCases_Rushby_2015}).
In this context, it is important to reason about the strengths and weaknesses of the evidence and to present this information clearly to stakeholders.

Another route is to analyze the system \emph{compositionally} instead of treating it as a black box~\cite{CompositionalReasoning_Giannakopoulou_18,Contracts_Benveniste_18,CPSFalsification_Dreossi_19}.
Consider a system being analyzed as a whole via simulation testing to acquire a probabilistic guarantee of correctness, a potentially expensive and time-consuming task.
The results could guide developers to make changes in order to improve system performance, but these changes invalidate the prior analysis.
If one instead analyzes the system compositionally by breaking it into components, only those that have changed need to be re-tested or verified. This is especially relevant for ML  components, which may undergo frequent changes from evolving training data or learning during deployment.

While many of the techniques above have been applied in isolation, to our knowledge there is no general theory that enables compositional analysis of CPS with black-box components operating in complex real-world environments while supporting sound, algorithmic combination of evidence ranging from explicit proofs to probabilistic testing into a formally-checked assurance case.

To address this gap, we propose a new tool, \textbf{\SPC{}}, built on top of the Scenic probabilistic programming language~\cite{Scenic_Fremont_19,Scenic_Fremont_22,Scenic3_Vin_23}, leveraging its ability to represent complex environments and test black-box systems. \textbf{\SPC{}} supports:
\begin{itemize}
    \item \textbf{Compositionally describing systems} with clear interfaces between system components and the outside world. Components can range from fully-interpretable code to opaque black boxes.
    \item \textbf{Assume-guarantee contracts for components}, using an extension of Linear Temporal Logic (LTL)~\cite{LTL_Pnueli_77} with predicates containing arbitrary Scenic expressions that can also compare values over time.
    \item \textbf{Testing in simulation} using Scenic to obtain probabilistic guarantees that a component satisfies a contract.
    \item \textbf{Formal proofs}, provided by a user, that a component satisfies a given contract or that a contract refines another. We interface to Lean~4~\cite{Lean4_Moura_21}, and our framework is general enough to add support for other tools in the future.
    \item \textbf{Using our underlying theory} to soundly combine the probabilistic results of verifying contracts while tracking what evidence supports which guarantees, culminating in the ability to generate a formally-checked assurance case for the entire system.
\end{itemize}

We also formalize the underlying theory used in ScenicProver, which contains:
\begin{itemize}
    \item \textbf{Abstract verification procedures} representing different ways of bounding the probability that a contract holds.
    \item \textbf{Judgment rules} which we prove can be used to soundly combine verification procedures using contract operators.
    \item \textbf{A novel weak merge operator} that can generate \emph{weaker contracts} with \emph{higher likelihood of correctness}.
\end{itemize}

In Section~\ref{sec:case_study}, we present two case studies of our framework applied to a simplified automatic emergency braking (AEB) system and a drone. We describe each inside \SPC{}, and use reasonable assumptions along with formal proofs to avoid testing a portion of the input space. These results are then combined to obtain a probabilistic guarantee that each system satisfies given top-level contracts. We obtain stronger results than testing the systems as monoliths, for the same computational budget. Our examples illustrate that by utilizing designer knowledge of how to split the space compositionally and leveraging known guarantees, one can significantly outperform existing techniques for verification of learning-enabled CPS while still ensuring a sound result.

The rest of the paper is structured as follows. In Section~\ref{sec:motivating_example} we introduce a motivating example that we use throughout the remainder of the paper. In Section~\ref{sec:preliminaries} we discuss existing concepts that our framework builds on. In Section~\ref{sec:contract_verification} we discuss the different methods of verifying contracts in our framework, leading to a theory of how to soundly combine those results in Section~\ref{sec:contract_combination}. Finally in Section~\ref{sec:case_study} we return to our motivating example as a case study, in addition to a drone-based case study, and demonstrate in both that using our framework provides a {\it stronger result faster} than what is achieved by testing each system as a monolith. Throughout the paper, proofs are omitted for space, and can be found in the Appendix.

\paragraph{Related Work.}
\label{sec:related_work}
Our work builds off existing theory and tools for verification of CPS, compositional verification, contract-based design, and assurance case generation.

Much existing work has applied testing/falsification-based techniques to CPS verification (see \cite{SurveyCPSValidation_Corso_2022} for a survey); almost all of these approaches, including Scenic~\cite{Scenic_Fremont_19} and VerifAI~\cite{VerifAI_Dreossi_19}, treat the system as a monolith. Furthermore, while falsification (using tools like VerifAI) is useful for finding counterexamples where a system violates a specification, it does not provide probabilistic guarantees about the behavior of a system given a specific environment model, which is the focus of our approach. Tools like KeYmaera X~\cite{KeYmaeraX_Fulton_15} and Verse~\cite{Verse_Li_23} allow the user to prove properties about CPS by modeling the controls and physical behavior of a system, but they have limited ability to represent complex real-world environments and learning-enabled components.

Other approaches to verification include runtime enforcement techniques which detect and correct invalid behavior, such as shields~\cite{Shields_Baier_2015} and the Simplex architecture~\cite{Simplex_Seto_98,Simplex_Sha_01}. There has been significant work on creating and synthesizing runtime monitors that enforce a given safety property~\cite{Soter_Desai_19,SynthesisRuntimeMonitors_Torfah_22,Ulgen_Yalcinkaya_23,BlackBoxRuntimeMonitors_Torfah_22}. These techniques 
are complementary to our approach, and can be used inside a system being verified by our framework (see Sec.~\ref{sec:case_study}).

Contract-based design~\cite{Contracts_Benveniste_18,2025algebraicaspectscontracts} is a framework that allows for compositional construction, abstraction, analysis, and verification of systems. In contract-based design, components define the behavior and data-flow of a system, and contracts define properties on individual components. Various tools, such as OCRA~\cite{OCRA_Cimatti_2013} and Pacti~\cite{pactipaper}, exist for reasoning about and manipulating contracts themselves, though they do not have the ability to reason about whether or not contracts hold over complex learning-enabled systems in realistic environments. Recent papers describe probabilistic contracts~\cite{TheoryProbabilisticContracts_Hampus_2024,10.1007/978-3-031-75380-0_3}, extending existing theory to the probabilistic case in a similar way to our formulation, but concern themselves primarily with the abstract theory and not the additional elements that our framework encompasses, such as methods of contract verification, proofs of contract refinement, heuristics for combining probabilistic contracts, etc.

Existing work has applied compositional analysis to falsification~\cite{CPSFalsification_Dreossi_19} and verification~\cite{CompositionalVerification_Pasareanu_18,ClosedLoopAnalysis_Pasareanu_23} of CPS, though not as a general framework that supports arbitrary system architectures. Existing work has also explored learning contracts over specific portions of the system~\cite{PerceptionContractsSafety_Astorga_2023}, or the assumptions for the whole system~\cite{IntrospectiveEnvironmentModeling_Seshia_2019}, but again these apply to specific types of contracts and situations (e.g., contracts over perception components or assumptions over the whole system).

Significant work exists on presenting and reasoning about the results of compositional verification in an assurance case~\cite{UsesArgument_Toulmin_2003,JustificationDiagram_Polacsek_2018,GoalStructuringNotation_Kelly_2004}, including tools for manipulating or generating aspects of them~\cite{EvidentialToolBus_Rushby_2005,AdvoCATE_Denney_12,ACGeneration_Wang_23,ContractACSynthesis_Wang_22}.
These are complementary to our work; importantly, they do not provide an algorithmic way to derive sound probability bounds when mixing testing, proofs, and probabilistic assumptions.

In summary, to our knowledge no framework has been proposed that supports compositional verification for CPS with black-box components over complex real-world environments, and no existing contract theory enables deriving probability bounds for contracts in a way that (1) supports all basic contract operators needed in practice, including our novel weak merge operator; (2) soundly incorporates probabilistic results from testing black-box components; and (3) allows effective, automatic computation of the probability bounds.

\section{Motivating Example}
\label{sec:motivating_example}
We motivate the utility of our framework with an example which we will use throughout the paper and in one of our case studies. While simplified, this example includes many elements of real-world systems, including closed-loop feedback, untrusted components, sensor fusion and robust state estimation for managing uncertainty, continuous signals, and complex dynamics.

Suppose that we are the manufacturers of an autonomous vehicle that is required by law to have an Automatic Emergency Braking (AEB) system, for which regulators have specified minimum standards: the vehicle must never get closer than $x$ meters to the car in front of it in at least $y$\% of cases with $z$ confidence. We must provide an argument to regulators for why our system meets this standard (such as an assurance case~\cite{AssuranceCases_Rushby_2015}) before our vehicle is allowed on the road.

Other manufacturers produce sensors that we plan to use in our system. These sensors have specifications which conform to their own sets of regulations, so we assume any such specification holds in our argument. In our example, we will utilize two such sensors whose error is guaranteed to be within a certain bound under certain conditions: a radar distance sensor which has guarantees if the car in front has at least a certain width, and a laser distance sensor which has guarantees if the atmosphere is free of occluding particles like rain and snow. Outside of these conditions the sensors should ideally still provide somewhat reasonable values, but the sensor manufacturers make no guarantees, so it is our responsibility to ensure the system meets regulator specifications across actual driving conditions. We also include a third sensor: a machine learning-based camera system which we believe will work well in practice but which has no guarantees.

\begin{figure}[tb]
    \centering
    \includegraphics[width=0.75\linewidth]{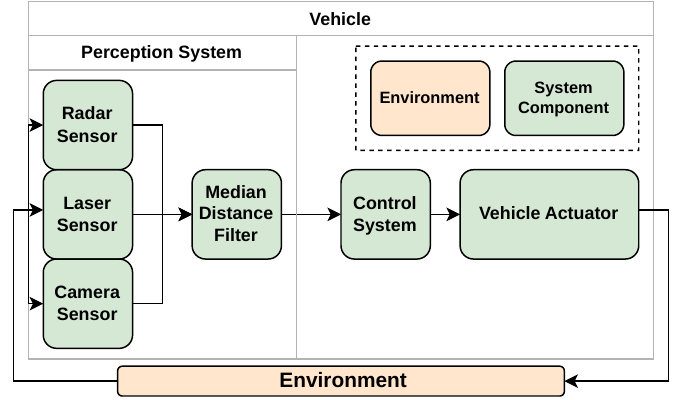}
    \caption{The structure of the system in our motivating example.}
    \label{fig:case_study}
\end{figure}

Testing our whole system in simulation provides a probabilistic guarantee of correctness; however, it does not allow us to make use of the guarantees provided by the sensor manufacturers, and means that any change to the system invalidates all previous tests. 
Using \SPC{}, we can do better.
There are three main steps:

\customparagraph{Components:}
We first split our system into discrete \textit{components}:
\begin{itemize}
    \item \textbf{Perception System:} Observes the world using sensors and outputs an estimate of the distance to the car in front.
    \item \textbf{Control System:} Inputs the distance estimate and outputs an appropriate throttle signal.
    \item \textbf{Vehicle Actuator:} Inputs the throttle signal and takes a corresponding action, either accelerating or braking.
\end{itemize}
Note that each component above might be made of other components. In fact, as a simple form of sensor fusion we will state that the perception system contains a component for each sensor listed above and a component which returns the median of their outputs (to reduce noise and errors). A visualization of this system can be found in Figure \ref{fig:case_study}.

\customparagraph{Contracts:}
With this structure in place, we can specify \textit{contracts} for each component in \SPC{}, which we want to hold over each component. Ideally these contracts, when combined using \SPC{}'s contract operators, should result in a contract representing the system-level specification we care about (i.e., not crashing into the car ahead of us). Suitable contracts for each of our components are:
\begin{itemize}
    \item \textbf{Perception System:} Outputs a distance accurate to within a given error bound.
    \item \textbf{Control System:} If the input distance is below a ``danger threshold'' depending on our current speed, the output signal should be to brake fully.
    \item \textbf{Vehicle Actuator:} If we brake fully, our speed decreases by a given amount.
\end{itemize}
When combining our contracts, we may need to provide an argument (via a machine-checkable proof) that they together imply that we satisfy our system-level specification of not crashing into the car ahead of us (i.e., the distance is relatively accurate, so the car brakes, so we slow down, so we do not hit the lead car).

\customparagraph{Verify:}
With the component structure of our system in place, and the contracts we want to apply to each component defined, we turn to actually showing the system correct. At this stage, we must \textit{verify} that these contracts hold, and we now have more options than when we viewed the system as a monolith.
We may be able to prove the contract on the control system outright, as it does not depend on the environment. Our contract on the vehicle actuator corresponds to braking deceleration, a value we may easily be able to test separately and import into our argument. We could also simply test our contract on the perception system, but there is a better approach that makes use of the guarantees we have been provided about the sensors in use. Namely, we could further subdivide our contract, splitting it into one contract that covers performance under the conditions our radar and laser sensors are known to work well in, and another contract covering all other conditions. The first contract we can prove, using our sensor assumptions and the semantics of the median filter, showing that our overall perception system will always be accurate in those conditions. Therefore, we can avoid testing this contract, and thus a fraction of the input space, and focus our limited testing budget on the other contract covering the ``unsafe'' part of the space. We do exactly this in Section~\ref{sec:case_study}, providing a stronger result than monolithic testing would be able to provide alone.

\section{Preliminaries: Traces, Components, and Contracts}
\label{sec:preliminaries}
In this section, we discuss existing concepts that form the backbone of our framework, and how we have implemented these concepts. These include how we model our system and the environment, and the ideas underpinning compositional verification including components and contracts.

\subsection{Environment and System Traces}
We model behaviors of the system and environment as \textit{traces}. A trace is a finite\footnote{In order to allow evaluating the satisfaction of specifications via testing, we consider only finite traces in this paper. We leave exploring techniques for obtaining infinite-time guarantees from finite testing plus additional assumptions for future work.} sequence of states of an environment and system (one state per discrete timestep).
A distribution over environment states is modeled by the random variable $\mathcal{E}$, which is defined by a Scenic program (we also refer to $\mathcal{E}$ as a \textit{scenario}). A specific environment state $e$ can be drawn from $\mathcal{E}$ with probability $\mathbb{P}(\mathcal{E} = e)$ (we also refer to $e$ as a \textit{scene}).

A component $\mathcal{M}$ (discussed in depth in Section~\ref{sec:components}) has a value $v$, defined to be the value of all IO and internal state at a given timestep.
A component value $v'$ can be retrieved by ``running'' a component: $v' = \mathcal{M}(e, v)$, where $v$ represents the previous state of $\mathcal{M}$ and can be either another component state or null (written $\emptyset$), which represents the deterministic initial state of a component. As we will see later, a component can be composed of other sub-components, and in this case we recursively define a component value to be the concatenation of the values of all sub-components.

To model how the environment changes over time in response to the system's output, we use an \textit{environment transition function}. Note here that the environment includes everything other than the components, and that the environment transition function is intentionally opaque so that it is able to model real-world environments for which our only assumption is that we can simulate them. The environment transition function $\mathcal{S}$ is formally defined as a function that takes an environment state $e$, a component value $v$, and a latent state $s$, returning a distribution over a pair containing a new environment state and latent state. Formally $(e', s') \leftarrow \mathcal{S}(e, v, s)$, where the notation $X \leftarrow D$ means that $X$ is a random variable sampled from distribution $D$. The distribution represents the probabilistic nature of $e$, in that it defines a distribution over future behaviors of environment agents. The latent state $s$ is used to represent unobserved world state.
The length of a trace is determined by the termination criteria in the Scenic program: we model the end of a trace by having $\mathcal{S}$ return a special environment value $\bot$.

Given a component $\mathcal{M}$, an environment state $e_0$, and a latent state $s_0$, we can sample a trace $\tau = ((e_0, v_0),\allowbreak \hdots,\allowbreak (e_m, v_m))$ by setting $v_0 = \mathcal{M}(e_0, \emptyset)$ and for all $i \ge 1$ until $e_{i-1} = \bot$, sampling $(e_i, s_i) \leftarrow \mathcal{S}(e_{i-1}, v_{i-1}, s_{i-1})$ and assigning $v_i = \mathcal{M}(e_i, v_{i-1})$.
We write $\mathcal{T}(e_0, s_0, \mathcal{M})$ for the induced \emph{trace distribution}, where $\tau$ has probability $\prod_{i=1}^{m}\mathbb{P}\left((e_{i}, s_{i}) = \mathcal{S}(e_{i-1}, v_{i-1}, s_{i-1}) \right)$.
We also overload $\mathcal{T}$ to be defined with regards to a scenario instead of a scene, writing $\tau \leftarrow \mathcal{T}(\mathcal{E}, s_{0}, \mathcal{M})$, where $\mathbb{P}(\mathcal{T}(\mathcal{E}, s_0, \mathcal{M}) = \tau) = \mathbb{P}(\mathcal{E} = e_0) \cdot \mathbb{P}(\mathcal{T}(e_0, s_{0}, \mathcal{M}) = \tau)$. As an abuse of notation, we write $\forall \ \tau \in \mathcal{T}$ to indicate quantifying over all possible traces that can be drawn from $\mathcal{T}$.

\subsection{Components}
\label{sec:components}
Components define a compartmentalized piece of a system, with well-defined inputs and outputs. As we will see in this section, components can be combined via \emph{composition}, such that the top-level system can also be considered a component.

The interface of a component is a 4-tuple $(I, O, S, A)$, where $I$ is a set of typed inputs (from the output of another component), $O$ is a set of typed outputs, $S$ is a set of typed sensor values (inputs from the environment), and $A$ is a set of typed actions (outputs to the environment).
Components can also have internal state.
All of these values together constitute the value of the component at a given timestep.

\begin{figure}[tb]
\centering
\includegraphics[width=\linewidth]{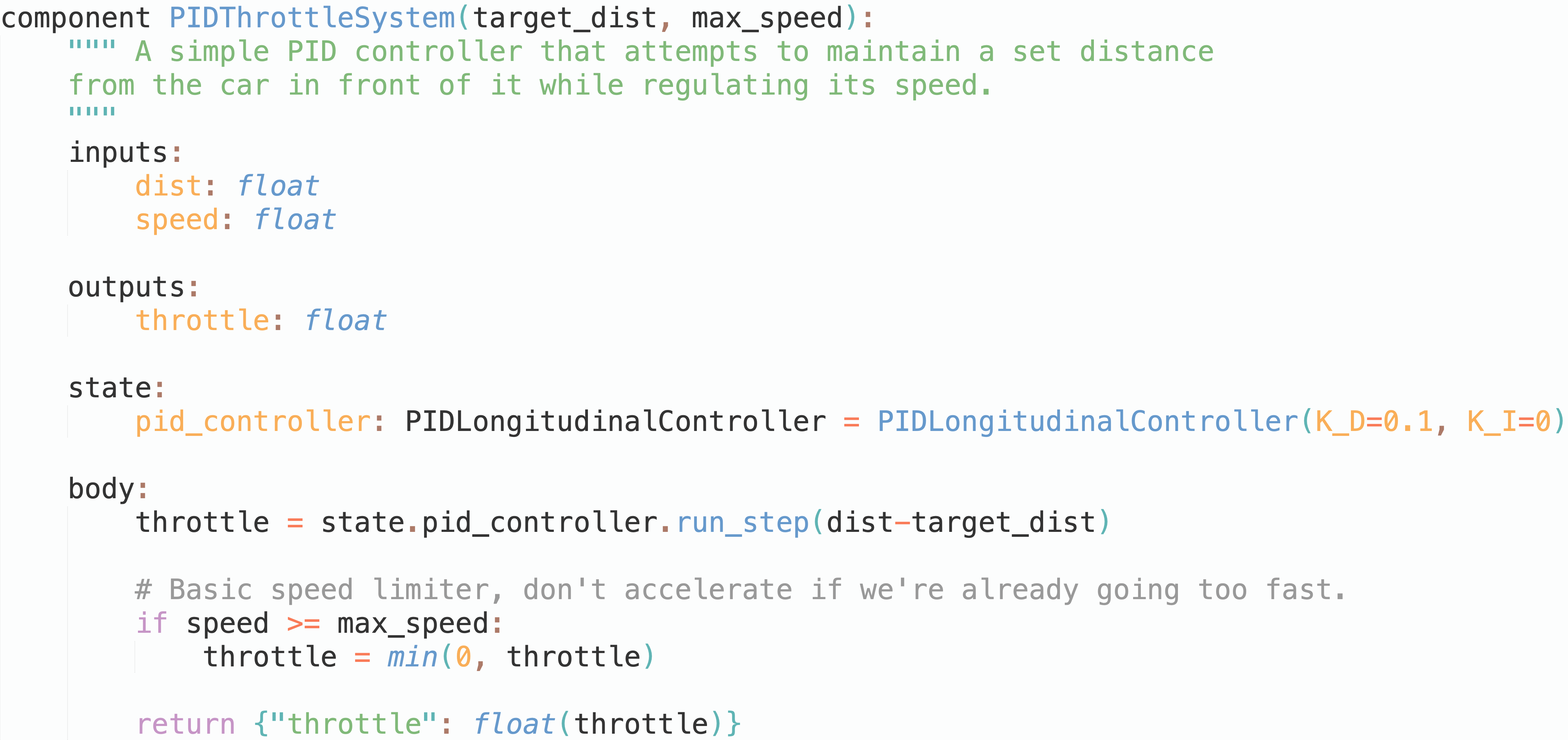}
\caption{A \SPC{} snippet defining the throttle controller component of the AEB example from Sec.~\ref{sec:motivating_example}.}
\label{fig:components_example_code}
\end{figure}

In Scenic, a system's implementation is defined by \emph{behavior} statements, which can contain arbitrary Python code, take arbitrary input from the world, and have arbitrary effects on the world. This approach provides significant flexibility to the user, while allowing black-box testing, but poses a problem for compositional reasoning. We take a different approach to defining components in our framework that balances expressivity with providing structure for analysis while preserving the ability to \textit{simulate any component}. Specifically, we add new syntax which restricts how components take input and affect world state to the defined component interfaces, which aids analysis.
The syntax defining the throttle controller component of our AEB example (part of the Control System module in Fig. 1) is shown in Figure~\ref{fig:components_example_code}.

\subsection{Contracts}
\begin{figure}[tb]
\centering
\includegraphics[width=\linewidth]{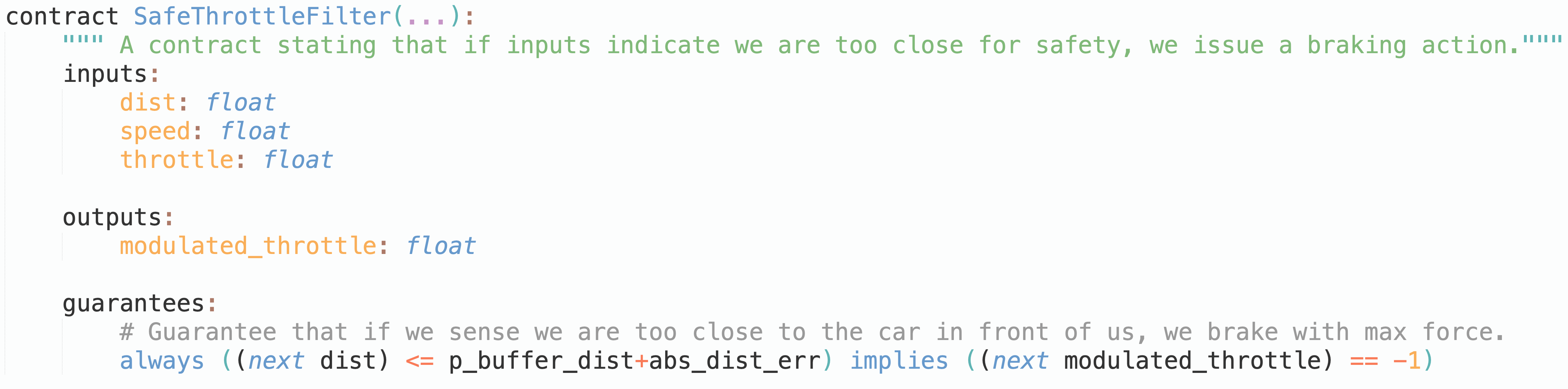}
\caption{An abbreviated \SPC{} snippet defining a contract over a component which overrides the throttle controller to apply emergency braking if the lead car is too close.}
\label{fig:contracts_example_code}
\end{figure}

To specify and reason about the behavior of components, we use \textit{assume-guarantee} contracts, which we define as follows.
A contract $\mathcal{C}$ is a pair $(\mathcal{A}, \mathcal{G})$ of LTLf Modulo Theories~\cite{LTLfMT_Geatti_16,LTLfMTDecidability_Geatti_23} formulas (an extension of Linear Temporal Logic (LTL)~\cite{LTL_Pnueli_77}), whose atomic formulas may use any Scenic expressions, including references to global state (such as positions of objects) and contract ports.
For example, the formula written \texttt{always self.speed <= (next self.speed)} in \SPC{} syntax states that the object in question never decelerates.
Traces as defined above include all the information necessary to give a truth value to such formulas.
A given trace $\tau$ \emph{satisfies} a contract $\mathcal{C} = (\mathcal{A}, \mathcal{G})$, written $\tau \vDash \mathcal{C}$, iff $\tau \vDash (\mathcal{A} \Rightarrow \mathcal{G})$. Given a distribution over traces $\mathcal{T}$, its probability of satisfying a contract $\mathcal{C}$ is $\mathbb{P}(\mathcal{T} \vDash \mathcal{C}) = \mathbb{P}(\tau \vDash \mathcal{C} \mid \tau \leftarrow \mathcal{T})=\sum_{\tau \in \mathcal{T}} \mathbb{P}(\tau) \mathbb{P}(\tau \vDash \mathcal{C})$.
Figure~\ref{fig:contracts_example_code} shows an example contract definition.

\section{Contract Verification}
\label{sec:contract_verification}
We now turn to methods to quantify to what extent contracts hold, which we call \textit{verification procedures}. These procedures allow us to formalize the different types of ``evidence'' supported by the framework, and allow us to define a calculus for combining them into an assurance case in the next section.

We define a verification procedure for a contract as a (possibly randomized) procedure which computes a lower bound for the probability that the contract holds:

\begin{definition}[Verification Procedures]
    A \emph{verification procedure} $\mathcal{V}(\mathcal{T}, c)$ is a probabilistic procedure which takes in a distribution over traces $\mathcal{T}$ and a confidence $c \in [0,1]$ and outputs $p \in [0,1]$.
    A verification procedure $\mathcal{V}$ is \emph{sound} with respect to a contract $\mathcal{C}$, written $\mathcal{V}(\mathcal{T}, c) \vdash \mathcal{C}$, if and only if
    \begin{align*}
         \mathbb{P}\left[\mathcal{V}(\mathcal{T}, c) \leq \mathbb{P}(\mathcal{T} \vDash \mathcal{C})\right] \geq c .
    \end{align*}
\end{definition}
Intuitively, soundness means that if the verification procedure is invoked many times, at least a $c$ fraction of the time it will output a correct lower bound on the true probability that the contract is satisfied. For simplicity, we omit the parameters of verification procedures when not needed.

In our framework, we assume that all evidence is derived independently. Intuitively, this means that our confidence in one piece of evidence does not affect our confidence in another piece of evidence. This concept is formalized in Assumption~\ref{assum:verification_independence}.
\begin{assum}
    \label{assum:verification_independence}
    All verification procedures are performed independently. Formally,
    for all verification procedures $\mathcal{V}_1$, $\mathcal{V}_2$, trace distributions $\mathcal{T}_1$, $\mathcal{T}_2$, and $c_1, c_2, p_1, p_2 \in [0, 1]$,
    \begin{align*}
    &\mathbb{P}\left[p_1 \leq \mathcal{V}_1(\mathcal{T}_1, c_1) \;\land\; p_2 \leq \mathcal{V}_2(\mathcal{T}_2, c_2)\right] \\
    &= \mathbb{P} [p_1 \leq \mathcal{V}_1(\mathcal{T}_1, c_1)] \cdot \mathbb{P}[p_2 \leq \mathcal{V}_2(\mathcal{T}_2, c_2)].
    \end{align*}
\end{assum}

\subsection{Testing-based Verification Procedures}
\label{sec:testing_verification}

The most accessible way to verify a potentially black-box system is testing, something that Scenic is already capable of doing. We begin by defining \textit{contract checking procedures}, which given a set of traces provide a lower bound on the number of traces in the set which satisfy a contract.

\begin{definition}[Contract Checking Procedure]
    A \emph{contract checking procedure} $\mathcal{K}$ takes a set of traces $\hat{\mathcal{T}}$ and returns a count $k$. It is \emph{sound} with respect to a contract $\mathcal{C}$, written $\mathcal{K}(\mathcal{\hat{T}}) \vdash \mathcal{C}$, iff $ \ \mathcal{K}(\hat{\mathcal{T}}) \leq |\{\tau \in \hat{\mathcal{T}} \mid \tau \vDash \mathcal{C} \}|$.
\end{definition}
For simplicity, we omit the parameters of contract checking procedures when not needed. A sound contract checking procedure naturally yields a sound verification procedure\footnote{The verification is sound for the simulated system; we leave addressing the sim-to-real gap to future work.}:

\begin{definition}[Testing-based Verification Procedures]
    The \emph{testing-based verification schema} $\mathcal{V}_T$ takes in a contract checking procedure $\mathcal{K}$ and the number of tests\footnote{ScenicProver generates a fixed number of tests (specified by the user or derived from a time budget). This avoids pitfalls with sequential approaches~\cite{ClopperPearsonSequential}.} to run $n \in \mathbb{N}$, returning a verification procedure as follows:
    \begin{algorithm}[H]
        \caption*{$\mathcal{V}_T(\mathcal{K}, n)(\mathcal{T}, c)$ Definition:}
        \begin{algorithmic}[1]
            \State $\tau_1 \sim \mathcal{T}, \hdots, \tau_n \sim \mathcal{T}$
            \State $\hat{\mathcal{T}} := \{ \tau_1, \hdots, \tau_n \}$
            \State $k := \mathcal{K}(\hat{\mathcal{T}})$
            \Statex $\triangleright \,$ Compute one-sided Clopper-Pearson confidence interval~\cite{ClopperPearsonInterval_Clopper_1934}
            \State $p := \texttt{ClopperPearsonInterval}(k, n, c)$
            \State \Return $p$
        \end{algorithmic}
    \end{algorithm}
\end{definition}

\begin{theorem}[Soundness of Testing-based Verification Procedures]
    \label{thm:testing_vp}
    If for all $\hat{\mathcal{T}}$ sampled IID\footnote{This assumption is satisfied by Scenic, which can sample traces IID.} from any $\mathcal{T}$, $\mathcal{K}(\hat{\mathcal{T}})$ is a sound contract checking procedure, then for any contract $\mathcal{C}$ and $n \in \mathbb{N}$, $\mathcal{V}_T(\mathcal{K}, n) \vdash \mathcal{C}$.
\end{theorem}

The first type of verification procedure available in \SPC{} is an instantiation of a testing-based verification procedure using the above schema. Recall that Scenic has the ability to specify environments, sample static scenes from them, and then run dynamic simulations with the help of a simulator. During a simulation, the assumptions and guarantees provided by a contract can be evaluated to determine whether or not they hold. Leveraging these two pieces, we can generate data using Scenic and then use it to see if a component satisfies a contract, allowing us to implement a \textit{sound contract checking procedure}. This technique provides a baseline method to generate evidence for components that may be difficult or impossible to reason about using other techniques.

There are several potential methods to actually implement the required contract checking procedure. The most straightforward method, implemented in \SPC{}, is to sample a scene from Scenic and run a dynamic simulation, with the linked Scenic object using the behavior defined by the components in question.
Using this method, each run provides evidence that can be used for all contracts on the components. Note that using this approach to test the system as a whole is equivalent to what is currently available using Scenic. We leave other methods to future work.

\subsection{Proof-based Verification Procedures}
\label{sec:proof_verification}
While testing can be generically applied and requires little user effort, the evidence generated is probabilistic and can take significant computation time to generate. For these reasons we also provide the ability to explicitly prove contracts correct.
The following proof-based verification schema supports an arbitrary machine-checkable proof format.

\begin{definition}[Proof-based Verification Procedures]
The proof-based verification schema $\mathcal{V}_{P}$ takes in a proof $r$, a proof checker $\mathcal{R}$, and a contract $\mathcal{C}$, returning a verification procedure. We indicate a proof being validated by the checker as $\mathcal{R}(r, \mathcal{C}) = \top$, which should occur only when $r$ is a valid proof that $\forall \ \tau, \tau \vDash (\mathcal{A} \Rightarrow \mathcal{G})$. Then we define
    \begin{align*}
        \mathcal{V}_P(r, \mathcal{R}, \mathcal{C})(\mathcal{T}, c) = 
        \begin{cases}
        1 & \mathcal{R}(r, \mathcal{C}) = \top\\
        0 & \text{otherwise}\\
        \end{cases}
    \end{align*}
\end{definition}
(Note that $\mathcal{T}$ and $c$ are unused, as the proof is independent of the distribution of traces and the desired confidence.)

\begin{theorem}
    \label{thm:proof_vp}
    If $\mathcal{R}$ is a \emph{sound checker}, i.e., it only accepts valid proofs, then $\mathcal{V}_P(r, \mathcal{R}, \mathcal{C})$ is a sound verification procedure.
\end{theorem}

In our implementation, we enable this functionality by formalizing semantics for a purely imperative loop-free fragment of Scenic in the proof assistant Lean~4~\cite{Lean4_Moura_21} and allowing the user to provide proofs that a given contract is correct. \SPC{} provides functionality to export assumptions and guarantees from our internal logic, along with the translated component and proof obligations, using the LeanLTL library~\cite{LeanLTL_Vin_25}, which provides structure for reasoning about LTLf Modulo Theories. \SPC{} checks the validity of the proof supplied by the user using the Lean~4 REPL~\cite{Lean4REPL}.

Providing semantics for all of Scenic in Lean is highly non-trivial since it can contain arbitrary Python. Limiting what can be written in Scenic components in general would be restrictive to the ability to represent real-life systems. Instead, for Lean-based proofs, we target a subset of Scenic containing arithmetic and several other useful functions like ceilings, floors, minimum, maximum, etc. and thus limit which components can be verified directly.\footnote{This subset has some overlap with established theories in LTLf Modulo Theories~\cite{LTLfMT_Geatti_16}. However, even in our simple example we encounter real arithmetic with ceilings and floors, an undecidable theory. Using an interactive theorem prover like Lean~4 sidesteps the problem of decidability.} However we show in our case studies that even a small subset of Scenic is enough to achieve interesting results.

\subsection{Assumption-based Verification Procedures}
\label{sec:assumptions_verification}
There are cases where the user may have reason for believing a contract to be true that does not support a formal proof and is not conducive to testing. A user may also want to complete the structure of their argument before deciding how to generate the requisite evidence. For such cases, we give users the ability to assume a contract is correct with a given probability and confidence. From a theory perspective, we model assumptions by simply assuming that a sound verification procedure has generated the assumed probability for the given confidence.

\section{Combining Verification Procedures}
\label{sec:contract_combination}
We have now seen how to formalize components and contracts, and verify that a component satisfies a contract.
The natural next question is how to translate component-level results to system-level results, a process we accomplish using contract operators: \emph{composition}, \emph{conjunction}, \emph{refinement}, \emph{strong merge}, and \emph{weak merge}. All of these operators, with the exception of weak merge, are existing operators from the contract literature~\cite{pactipaper,2025algebraicaspectscontracts}. For each  operator, we provide a judgment rule deriving a sound verification procedure for the output of the operator, given sound procedures for its inputs. We provide additional details for the weak merge operator, a novel operator proposed here for the purpose of reducing manual proof burden by weakening the needed contracts.

\SPC{} computes the required verification procedures directly, using Pacti~\cite{pactipaper} to compute the outputs of many of the contract operators.
As checking refinement can be undecidable~\cite{LTLfMT_Geatti_16,LeanLTL_Vin_25}, we also provide the ability for the user to input a Lean~4 proof using LeanLTL~\cite{LeanLTL_Vin_25}
if Pacti is unable to prove refinement automatically.

\subsection{Refinement}
The \textit{refinement} operator provides a partial ordering over contracts. It is useful for translating a ``stronger'' contract into a ``weaker'' contract that has a desired form.

\begin{definition}[Refinement Contract Operator~\cite{pactipaper,2025algebraicaspectscontracts}]
    We say $\mathcal{C}_1 = (\mathcal{A}_1, \mathcal{G}_1)$ \emph{refines} $\mathcal{C}_2 = (\mathcal{A}_2, \mathcal{G}_2)$, i.e. $\mathcal{C}_1 \leq \mathcal{C}_2$, iff
    $\forall \tau, \tau \vDash (\mathcal{A}_2 \Rightarrow \mathcal{A}_1)  \land ((\mathcal{A}_1 \Rightarrow \mathcal{G}_1) \Rightarrow (\mathcal{A}_2 \Rightarrow \mathcal{G}_2))$.
\end{definition}

\begin{theorem}[Refinement Rule]
    \label{thm:refinement_vp}
    \begin{align*}
        \inferrule{
                    \mathcal{V}_1(\mathcal{T}, c) \vdash \mathcal{C}_1 \quad 
                    \mathcal{C}_1 \leq \mathcal{C}_2
                  }
                  {
                    \mathcal{V}_1(\mathcal{T}, c) \vdash \mathcal{C}_2
                  }
    \end{align*}
\end{theorem}

\subsection{Composition, Conjunction, and Strong Merge}
Composition combines contracts over sub-components to obtain a contract over the resulting higher-level component.
Conjunction and strong merge combine contracts over the \emph{same} component, the first losslessly and the second providing guarantees only when both contracts apply.
They are defined as follows (see \cite{2025algebraicaspectscontracts} for more information):
\begin{definition}[Composition, Conjunction, and Strong Merge Operators~\cite{pactipaper,2025algebraicaspectscontracts}]
    \label{def:compose_conjunction_strongmerge_defs}\\
    {
    \centering
    \setlength{\tabcolsep}{6pt}
    \begin{tabular}{|c|c|c|}
    \hline
    Operator Name & Symbol    & \makecell{Assumptions \\ Guarantees} \\ \hline
    Composition   & $\parallel$ & \makecell{$(\mathcal{A}_1 \land \mathcal{A}_2) \lor (\mathcal{A}_1 \land \mathcal{G}_1) \lor (\mathcal{A}_2 \land \mathcal{G}_2)$ \\ $((\mathcal{A}_1 \Rightarrow \mathcal{G}_1) \land (\mathcal{A}_2 \Rightarrow \mathcal{G}_2)) \lor (\lnot \mathcal{A}_1 \land \lnot \mathcal{A}_2)$} \\ \hline
    Conjunction   & $\land$     & \makecell{$\mathcal{A}_1 \lor \mathcal{A}_2$ \\ $((\mathcal{A}_1 \Rightarrow \mathcal{G}_1) \land (\mathcal{A}_2 \Rightarrow \mathcal{G}_2)) \lor (\lnot \mathcal{A}_1 \land \lnot \mathcal{A}_2)$} \\ \hline
    Strong Merge  & $\bullet$   & \makecell{$\mathcal{A}_1 \land \mathcal{A}_2$ \\ $(\mathcal{G}_1 \land \mathcal{G}_2) \lor \lnot \mathcal{A}_1 \lor \lnot \mathcal{A}_2$} \\ \hline
    \end{tabular}\par}
\end{definition}

Our judgment rules for these operators derive from a common scheme $\mathcal{V}_{\cup}(\mathcal{V}_1, \mathcal{V}_2, *)$ which combines verification procedures $\mathcal{V}_1$ and $\mathcal{V}_2$ using the union bound:

\begin{theorem}[Composition, Conjunction, Strong Merge Rules]
\label{def:compose_conjunction_strongmerge_vp}\\
\begin{minipage}{0.48\textwidth}
    \begin{gather*}
        (1) \ \inferrule{
                    \mathcal{V}_1(\mathcal{T}, c_1) \vdash \mathcal{C}_1 \quad
                    \mathcal{V}_2(\mathcal{T}, c_2) \vdash \mathcal{C}_2
                  }
                  {
                    \mathcal{V}_{\cup}(\mathcal{V}_1, \mathcal{V}_2, \parallel)(\mathcal{T}, c_1 c_2) \vdash (\mathcal{C}_1 \parallel \mathcal{C}_2)
                  } \\
        (2) \ \inferrule{
                    \mathcal{V}_1(\mathcal{T}, c_1) \vdash \mathcal{C}_1 \quad
                    \mathcal{V}_2(\mathcal{T}, c_2) \vdash \mathcal{C}_2
                  }
                  {
                    \mathcal{V}_{\cup}(\mathcal{V}_1, \mathcal{V}_2, \land)(\mathcal{T}, c_1 c_2) \vdash (\mathcal{C}_1 \land \mathcal{C}_2)
                  } \\
        (3) \ \inferrule{
                    \mathcal{V}_1(\mathcal{T}, c_1) \vdash \mathcal{C}_1 \quad
                    \mathcal{V}_2(\mathcal{T}, c_2) \vdash \mathcal{C}_2
                  }
                  {
                    \mathcal{V}_{\cup}(\mathcal{V}_1, \mathcal{V}_2, \bullet)(\mathcal{T}, c_1 c_2) \vdash (\mathcal{C}_1 \bullet \mathcal{C}_2)
                  }
    \end{gather*}
\end{minipage}
\hfill
\begin{minipage}{0.4\textwidth}
    \begin{algorithm}[H]
    \caption*{$\mathcal{V}_{\cup}(\mathcal{V}_1, \mathcal{V}_2, *)(\mathcal{C}_1 * \mathcal{C}_2, \mathcal{T}, c_1 c_2)$:}
    \begin{algorithmic}[1]
        \State $p_1 \leftarrow \mathcal{V}_1(\mathcal{T}, c_1)$
        \State $p_2 \leftarrow \mathcal{V}_2(\mathcal{T}, c_2)$
        \State \Return $p_1 + p_2 - 1$
    \end{algorithmic}
\end{algorithm}
\end{minipage}
\end{theorem}

\subsection{Weak Merge}
The \emph{weak merge} operator returns a contract that assumes either of the assumptions and either of the  guarantees of the input contracts hold. This contract operator is novel to our work, because it produces a contract that is more relaxed in the refinement order than the result of conjunction or strong merging. However, it has useful properties when it comes to verification procedure combination. Intuitively, if the contract resulting from the weak merge operation is still strong enough for our needs, we can obtain a higher probability of correctness than using the conjunction or strong merge operator, as we need only assume that \emph{one} input assumption holds.

\begin{definition}[Weak Merge Contract Operator]
    Given two contracts $\mathcal{C}_1 = (\mathcal{A}_1, \mathcal{G}_1)$ and $\mathcal{C}_2 = (\mathcal{A}_2, \mathcal{G}_2)$, their \emph{weak merge} $\mathcal{C}_1 \bowtie \mathcal{C}_2$ is $(\mathcal{A}_1 \lor \mathcal{A}_2, \; \mathcal{G}_1 \lor \mathcal{G}_2 \lor \lnot (\mathcal{A}_1 \lor \mathcal{A}_2))$.
\end{definition}

\begin{theorem}[Weak Merge Rule]
    \label{thm:weak_merge_vp}
    \begin{align*}
        \inferrule{
                    \mathcal{V}_1(\mathcal{T} \mid \mathcal{A}_1, c_1) \vdash \mathcal{C}_1 \quad
                    \mathcal{V}_2(\mathcal{T} \mid \lnot \mathcal{A}_1, c_2) \vdash \mathcal{C}_2
                  }
                  {
                    \mathcal{V}_{\bowtie}(\mathcal{T}, c_1 c_2) \vdash (\mathcal{C}_1 \bowtie \mathcal{C}_2)
                  }
    \end{align*}
    \begin{algorithm}[H]
    \caption*{$\mathcal{V}_{\bowtie}(\mathcal{V}_1, \mathcal{V}_2)(\mathcal{C}_1 \bowtie \mathcal{C}_2, \mathcal{T}, c_1 c_2)$ Definition:}
        \begin{algorithmic}[1]
            \State $p_1 \leftarrow \mathcal{V}_1(\mathcal{T} \mid \mathcal{A}_1, c_1)$
            \State $p_2 \leftarrow \mathcal{V}_2(\mathcal{T} \mid \lnot \mathcal{A}_1, c_2)$
            \State \Return $p_1 \cdot \mathbb{P}(\mathcal{T} \vDash \mathcal{A}_1) + p_2 \cdot (1 - \mathbb{P}(\mathcal{T} \vDash \mathcal{A}_1))$
        \end{algorithmic}
    \end{algorithm}
\end{theorem}

As this verification procedure requires knowing $\mathbb{P}(\mathcal{T} \vDash \mathcal{A}_1)$, \SPC{} implements a less general but more practical procedure for the case where we have a proof for one contract (checked using $\mathcal{V}_P$ from Sec.~\ref{sec:proof_verification}).
We present it as a sound contract checking procedure $K_{\bowtie T}$, which can be converted to a sound verification procedure.

\begin{theorem}[Weak Merge Contract Checking Procedure]
    \label{thm:weak_merge_testing_vp}
    \begin{align*}
        \inferrule{
                    \mathcal{V}_P(r, \mathcal{R}, \mathcal{C}_1)(\mathcal{T}, 1) \vdash \mathcal{C}_1 \quad
                    \mathcal{K}_{1}(\hat{\mathcal{T}_T}) \vdash \mathcal{C}_2
                  }
                  {
                    \mathcal{K}_{\bowtie T}(r, \mathcal{R}, \mathcal{K}_1, \mathcal{T}, n)(\mathcal{\hat{T}}) \vdash (\mathcal{C}_1 \bowtie \mathcal{C}_2)
                  }
    \end{align*}
    \begin{algorithm}[H]
        \caption*{$\mathcal{K}_{\bowtie T}(r, \mathcal{R}, \mathcal{K}_1, \mathcal{T}, n)(\mathcal{\hat{T}})$ Definition:}
        \begin{algorithmic}[1]
            \State $p' \leftarrow \mathcal{V}_P(r, \mathcal{R}, \mathcal{C}_1)(\mathcal{T}, 1)$
            \If{$p' < 1$} 
                \Return 0 \Comment{Proof of $\mathcal{C}_1$ did not check}
            \EndIf
            \State $\hat{\mathcal{T}_P} := \{ \tau \in \hat{\mathcal{T}} \mid \tau \vDash \mathcal{A}_1\}$ \Comment{$\mathcal{C}_1$ applies; no need to test $\mathcal{C}_2$}
            \State $\hat{\mathcal{T}_T} := \hat{\mathcal{T}} - \hat{\mathcal{T}_P}$ \Comment{Test $\mathcal{C}_2$ on the remaining traces}
            \State $k := \mathcal{K}_{1}(\hat{\mathcal{T}_T}, c)$
            \State \Return $k + |\hat{\mathcal{T}_P}|$
        \end{algorithmic}
    \end{algorithm}
\end{theorem}

As we will demonstrate in Sec.~\ref{sec:case_study}, this procedure provides a speedup over strong merge when an assumption can be checked \emph{statically}, i.e., traces with the same initial state will either all satisfy or all violate the assumption.
For example, suppose $\mathcal{C}_1$ assumes $\mathcal{A}$ and sunny weather, while $\mathcal{C}_2$ assumes $\mathcal{A}$ and rainy weather.
If we prove $\mathcal{C}_1$, and can determine that the weather is sunny in trace $\tau$ from its initial state, then we know that $\tau \vDash \mathcal{C}_1 \bowtie \mathcal{C}_2$ without needing to run a complete simulation.
Note that it is sufficient to be able to check the weather statically, regardless of the complexity of $\mathcal{A}$.
In general, for assumptions $\mathcal{A}_1$ and $\mathcal{A}_2$ which are conjunctions of constraints, only the constraints which are not common to both assumptions need to be checkable statically.
This is because if a constraint is included in both sets of assumptions, then if it holds, the proof of $\mathcal{C}_1$ ensures $\mathcal{C}_1 \bowtie \mathcal{C}_2$ holds; if instead the constraint does not hold, then $\mathcal{A}_1 \lor \mathcal{A}_2$ is violated and so $\mathcal{C}_1 \bowtie \mathcal{C}_2$ is vacuously true.

\subsection{Visualizing Verification Procedure Trees}
\SPC{} can visualize all generated contract results as a textual assurance case, showing a tree of the rules that generated the top-level result, down to the proofs, assumptions, and tests that were used in the base verification procedures.
Complete examples are shown in Appendix~\ref{appendix:example_assurance_cases}.
In the future, we hope to add support for other tools/methods~\cite{UsesArgument_Toulmin_2003,JustificationDiagram_Polacsek_2018,GoalStructuringNotation_Kelly_2004,AdvoCATE_Denney_12} for visualizing assurance cases.

\section{Case Studies}
\label{sec:case_study}
We demonstrate the applicability and utility of our framework with case studies on our motivating AEB example and on an additional drone example. The AEB example uses Lean proofs written entirely by hand, which took significant time, while the drone example proofs were written using Claude Code~\cite{ClaudeCode} in only a few minutes, despite both case studies being of similar complexity. It is important to note that ScenicProver generates the theorem statements directly from the contracts, and checks the proofs using the Lean~4 REPL~\cite{Lean4REPL}.
This allows modern AI-based theorem provers to accelerate the use of ScenicProver without compromising correctness.
Both case studies are somewhat simplified in order to better illustrate the key features of our framework, leaving a larger industrial case study for future work.
In both case studies, we use a combination of proofs and testing to get a stronger result than testing the system as a monolith could provide.
Full code for the components, contracts, and proofs is available in our repository (Figures~\ref{fig:components_example_code} and \ref{fig:contracts_example_code} give abbreviated examples from the AEB case study).

\subsection{AEB Example}
\customparagraph{Components:}
We begin by laying out the components of our case study in our \texttt{.contracts} file. Most of the case study components are as described in Sec.~\ref{sec:motivating_example}. For simplicity, we model the three distance sensors as components that take the true distance to the lead vehicle and add errors to simulate their real-world analogues. The radar sensor has a chance of failure (returning a too-high value) that increases as the profile of the lead car shrinks, simulating the radar profile of the lead vehicle being too small to reliably detect. The laser sensor has a chance of failure (returning a too-small value) in conditions like rain and snow, where atmospheric particles might obstruct the beam. The camera sensor simply adds a small amount of Gaussian noise to the true value.

\begin{figure*}[tb]
    \centering
    \includegraphics[width=\linewidth]{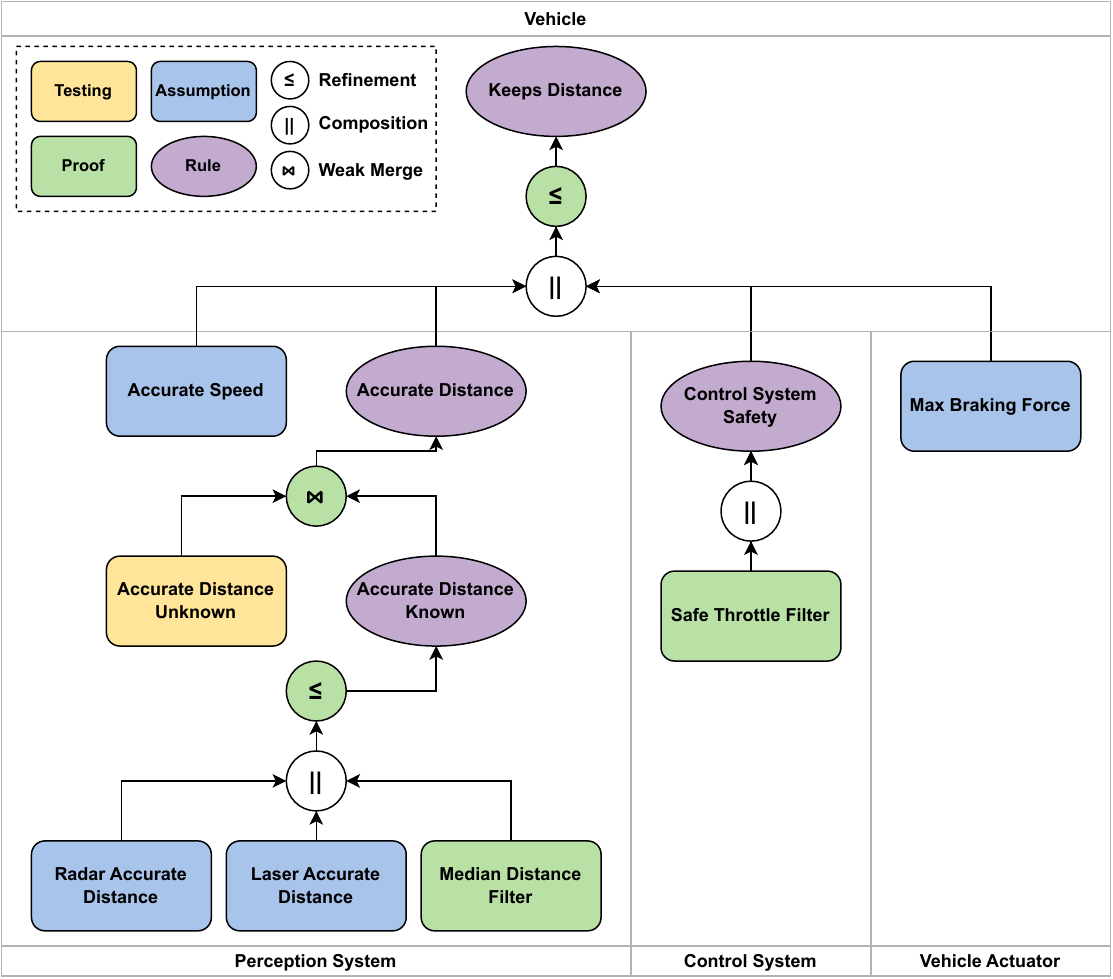}
    \caption{The structure of the AEB assurance case.}
    \label{fig:proof_tree}
\end{figure*}

\customparagraph{Contracts:}
We now turn to constructing the skeleton of our assurance case in \SPC{}, which is visualized in Figure~\ref{fig:proof_tree}. We will reach our overall system safety contract (\textsc{Keeps Distance}) by refining the composition of contracts over the three high-level components of our vehicle: the perception system, control system, and vehicle actuator. 
For the perception system, we compose and refine the three sensor contracts linked to our manufacturer guarantees into a contract over the whole perception system that shows that when the assumptions of the radar and laser sensors are met, the perception system will output an accurate distance (\textsc{Accurate Distance Known}). We then define another contract over the perception system covering the remaining portion of the space (\textsc{Accurate Distance Unknown}) and perform a weak merge over these two contracts, resulting in our overall \textsc{Accurate Distance} contract. We also state a contract guaranteeing an accurate speed reading.
For the control system and vehicle actuator, our contracts mimic those described in Section~\ref{sec:motivating_example}: namely, we will always output a braking action if we are perceived to be too close to the vehicle ahead of us, and if we apply a maximal braking command, we slow down by a given amount.
The operations shown in the figure are those explicitly invoked in the \SPC{} file, though \SPC{} syntax does abstract away some boilerplate.\footnote{For example, the composition to derive \textsc{Control System Safety} is over only the \textsc{Safe Throttle Filter}, using an implicit vacuous contract for the PID controller.} All operators with a white background were computed automatically by Pacti, while refinements with a green background were proved manually in Lean 4. Some of these proofs were highly non-trivial (e.g. the proof of refinement for \textsc{Keeps Distance} uses an undecidable theory: nonlinear arithmetic with ceilings and floors).

\customparagraph{Verify:}
We must now instruct \SPC{} as to what evidence it should use in the assurance case. We begin with the contracts for the control system and vehicle actuator. For the control system, we prove in Lean 4 that the safety filter (i.e. a shield) will always output a braking action if we are perceived to be too close to the vehicle ahead of us. This guarantee is transferred to the control system as a whole automatically via Pacti when the control system is composed together. For the vehicle actuator, we assume that a maximal braking action will result in a given slowdown with a probability of at least 99\% with 99.9\% confidence, with the justification used in Section~\ref{sec:motivating_example} that this could be determined by outside testing.

Turning to the perception system, \textsc{Accurate Speed} is assumed directly (justified by assuming an accurate speedometer). For showing our \textsc{Accurate Distance} contract, we begin by importing our manufacturer sensor requirements (\textsc{Radar Accurate Distance} and \textsc{Laser Accurate Distance}) and proving a small contract about the median filter (\textsc{Median Distance Filter}). This flows through our assurance case and gives us a guarantee on \textsc{Accurate Distance Known}, leaving \textsc{Accurate Distance Unknown} which we will test via simulation with a target confidence of 99.9\%. The end result is that \textsc{Accurate Distance Known} (which represents 35\% of the space in our example) need not be tested, and so \SPC{} is able to allocate all of our testing budget to the remaining 65\% of the space covered by \textsc{Accurate Distance Unknown}.

\SPC{} propagates the evidence from each subsystem to the top of the assurance case, and provides an overall probabilistic guarantee on \textsc{Keeps Distance}.

\begin{figure*}[tb]
    \centering
    \begin{subfigure}[t]{0.49\textwidth}
        \includegraphics[width=\textwidth]{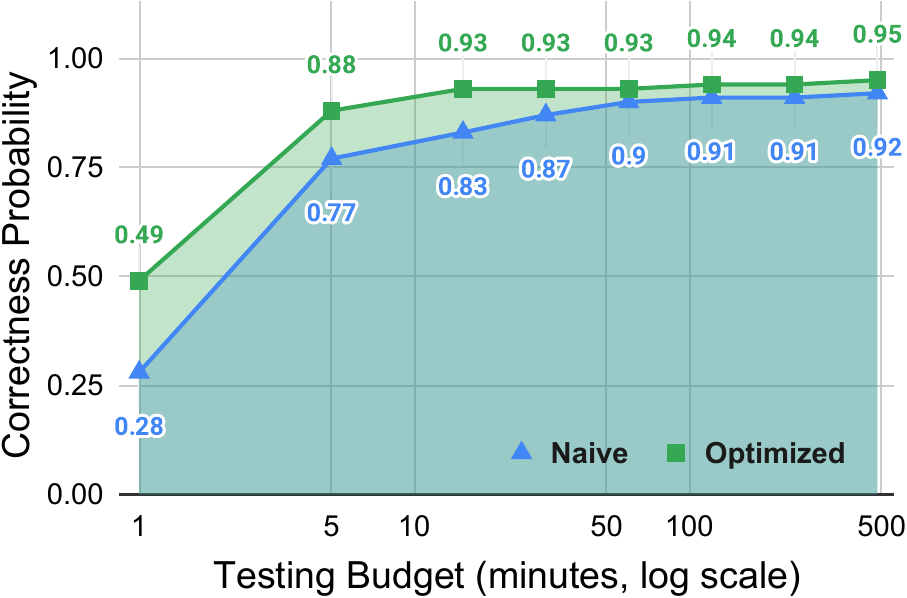}
        \caption{AEB Contract (\textsc{Keeps Distance})}
        \label{fig:contract_probability_time}
    \end{subfigure}
    \begin{subfigure}[t]{0.49\textwidth}
        \centering
        \includegraphics[width=\textwidth]{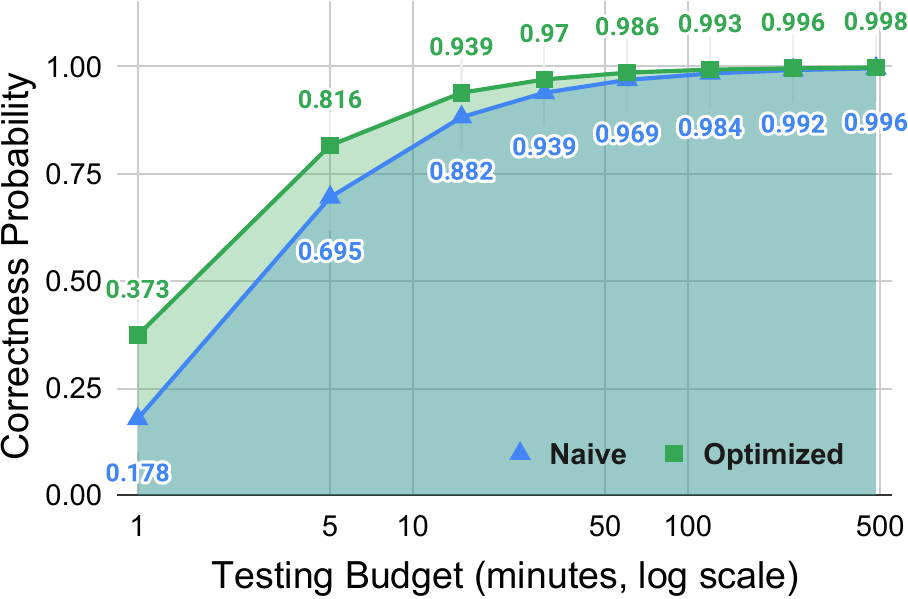}
        \caption{Drone Contract (\textsc{Corridor Containment})}
        \label{fig:drone_probability_time}
     \end{subfigure}
    \caption{Probability lower bounds for the top-level case study contracts, as a function of time spent testing.}
\end{figure*}

The gains achieved by structuring the assurance argument in this way rather than testing the system as a monolith are shown in Figure~\ref{fig:contract_probability_time}.
We compare the probability bounds obtained for the top-level contract between using \SPC{} na\"ively versus incorporating the manufacturer guarantees using the weak merge operator.
The results are shown as a function of time spent testing, up to a total of 8 hours.
As can be seen in the figure, the optimized analysis, possible only because of the reasoning abilities of \SPC{}, results in us being able to show our overall guarantee is true with probability 94.59\%, up from 91.55\% with the same time budget and confidence (99.8\%).
In fact, {\it \SPC{} is able to provide a stronger result in 15 minutes than what na\"ively testing the system as a monolith can provide in 8 hours}.
The complete assurance cases generated by \SPC{} after 8 hours (na\"ive and optimized) are given in  Appendix~\ref{appendix:example_assurance_cases}.

\subsection{Drone Example}

Our second case study is a quadrotor drone flying in wind, simulated in AirSim \cite{AirSim}. The drone must fly along a wall, maintaining a desired distance to it and the ground.  

\customparagraph{Components:}
We model two position estimators, GNSS and visual-inertial odometry (VIO), as components that take the drone's true position from the simulator and add errors characteristic of each sensor type. With ``good satellite reception,'' GNSS readings have only small, bounded Gaussian noise; with ``poor reception,'' an occasional reading can be off by a much larger margin. Reception quality is a random parameter sampled at the beginning of each simulated episode, while measurement errors are sampled at each timestep. Wind may degrade both the drone's perception and its ability to actuate accurately: it corrupts the drone's position estimates and limits how faithfully the drone tracks its velocity commands. In particular, VIO behaves well in light wind, but as stronger wind tilts the drone, it loses track of the ground features it was tracking and its errors grow larger and more frequent. Both sensors' estimates are converted into distances to the wall and the floor. For each surface, a median filter combines the two sensor readings with the previous fused value, so that an isolated bad reading from either sensor has limited effect. The fused distances are read by controllers which steer the drone toward a target clearance from each surface without exceeding a maximum approach rate. Appendix~\ref{appendix:drone_parameters} lists the numerical values of the parameters of the components.

\customparagraph{Contracts:}
Figure~\ref{fig:drone_assurance_case} in the Appendix shows the structure of the drone's assurance case. At its root is \textsc{Corridor Containment}, which asserts a safety property and a progress property: {\it the drone must never come within a minimum clearance of the wall or the floor, but whenever it is far enough from either of the surfaces, it must approach it}.

On the perception side, each sensor's contract bounds the error between the distance reported by the sensor and the true distance, but only under that sensor's rated conditions: adequate satellite reception for GNSS and low horizontal wind for VIO. The median filter's contract carries this bound through fusion, so that whenever both sensors are within their rated conditions, the fused estimate obeys the same bound.

On the actuation side, the contract limits how much the drone's actual per-timestep displacement toward each surface can differ from the commanded velocity, with slack for the lag in responding to new commands. This bound is what supports the progress half of \textsc{Corridor Containment}: together with the controllers' commanded approach rates, it ensures the drone keeps moving toward a surface that is still far enough away.

Together with the controller contracts, these compose into a contract over the whole drone stating that \textsc{Corridor Containment} holds in every scene where all of the rated conditions are satisfied. A second contract covers the scenes where at least one condition fails, and a weak merge of the two yields the top-level contract.

\customparagraph{Verify:}
The controller and median-filter contracts are proved in Lean. The sensor contracts are assumed directly, standing in for the accuracy specifications a manufacturer might provide for each sensor under its rated conditions. The actuation contract is likewise assumed, on the grounds that such bounds could be established by characterizing the airframe's response to commands in wind, for example through wind-tunnel testing.

From this evidence, \SPC{} establishes the top-level contract on every scene inside the rated conditions without any simulation. Only the complementary contract, covering scenes outside those conditions, needs evidence from testing, which we gather by simulating the whole drone as it moves through the environment specified by the \texttt{.scenic} file. Because 53\% of the scene distribution lies inside all rated conditions, the entire simulation budget can be spent on the remaining 47\%.

Figure~\ref{fig:drone_probability_time} plots the resulting lower bound on \textsc{Corridor Containment} against simulation time, for both the compositional analysis and a monolithic testing baseline. After 8 hours, the compositional bound reaches 99.81\%, compared with 99.61\% for the baseline, both at 99.9\% confidence. At 4 hours, the compositional analysis already certifies 99.64\% correctness, more than the baseline achieves with the full 8 hours.

\section{Conclusion}
In this paper we introduced \SPC{}, a new framework that builds on the Scenic probabilistic programming language to allow for compositional verification of cyber-physical systems with learning-enabled components using a variety of methods from explicit proofs to testing. We illustrated the value of the system by providing a stronger verification result for the same testing budget on two example systems given reasonable assumptions, something that would not have been possible in Scenic alone.

In future work we plan to expand on this framework, adding additional methods of evidence generation such as proof-producing SMT solvers and analysis techniques for neural networks, further integration with Pacti~\cite{pactipaper} to automate more complex contract operations, techniques to address the sim-to-real gap, and support for synthesis of contracts from top-level requirements. We also plan to apply this framework to a real system in a more in-depth case study paper.

\begin{credits}

\end{credits}
%
%
%
\bibliographystyle{splncs04}
\bibliography{refs,additional_refs}
\pagebreak
\appendix

\section{Proofs of Base Verification Procedures}
\label{appendix:base_vp}
\textbf{Theorem~\ref{thm:testing_vp} Proof:}
\begin{proof}
    Given that $\mathcal{K}$ is a sound contract checking procedure, we know that $k$ is a lower bound on the true number of traces in $\hat{\mathcal{T}}$ that satisfy $\mathcal{C}$. Since $\hat{\mathcal{T}}$ is sampled IID from $\mathcal{T}$, the rest of the proof follows from the validity of the Clopper-Pearson interval~\cite{ClopperPearsonInterval_Clopper_1934}, an exact method for finding a confidence interval over a binomial distribution.
\end{proof}

\noindent\textbf{Theorem~\ref{thm:proof_vp} Proof:}
\begin{proof}~\\
    \textbf{Case} ($\mathcal{R}(r, \mathcal{C}) = \top)$:\\
    In this case, we know that $\mathcal{C}$ is valid as $\forall \ \mathcal{T}, \ \forall \ \tau \in \mathcal{T}, \ \tau \vDash (\mathcal{A} \Rightarrow \mathcal{G})$, so $\mathbb{P}(\mathcal{T} \vDash \mathcal{C}) = 1$. As $p$ is always 1 in this case, and $1 \leq 1$, $\mathcal{V}_P$ is sound in this case.\\
    \textbf{Case:} ($otherwise$)\\
    In this case, $\mathcal{V}_P(r, \mathcal{R}, \mathcal{C})(\mathcal{T}, c) = 0$. By definition, $\mathbb{P}(\mathcal{T} \vDash \mathcal{C}) \geq 0$, so $\mathcal{V}_P$ is trivially sound in this case.\\
\end{proof}

\section{Proofs of Contract Operator Verification Procedures}
\label{appendix:contract_operator_vp}
We begin by establishing two lemmas that follow from the existing contract literature.

\begin{lemma}
    \label{lemma:composition_prob}
    Given two contracts $\mathcal{C}_1$ and $\mathcal{C}_2$, and a trace distribution $\mathcal{T}$,
    \begin{align*}
        \mathbb{P}(\mathcal{T} \vDash \mathcal{C}_1 \parallel \mathcal{C}_2)  &= \mathbb{P}((\mathcal{T} \vDash \mathcal{C}_1) \land (\mathcal{T} \vDash \mathcal{C}_2)) \\        
    \end{align*}
\end{lemma}
\begin{proof}
    Follows from \cite{pactipaper,Contracts_Benveniste_18}.
\end{proof}

\begin{lemma}
    \label{lemma:conjunction_prob}
    Given two contracts $\mathcal{C}_1$ and $\mathcal{C}_2$, and a trace distribution $\mathcal{T}$,
    \begin{align*}
        \mathbb{P}(\mathcal{T} \vDash \mathcal{C}_1 \land \mathcal{C}_2)  &= \mathbb{P}((\mathcal{T} \vDash \mathcal{C}_1) \land (\mathcal{T} \vDash \mathcal{C}_2)) \\
    \end{align*}
\end{lemma}
\begin{proof}
    Follows from \cite{pactipaper,Contracts_Benveniste_18}.
\end{proof}

We now prove the soundness of the refinement, composition, conjunction, and strong merge verification procedures.

\noindent \textbf{Theorem~\ref{thm:refinement_vp} Proof:}
\begin{proof}
    As $\mathcal{V}_1$ is sound, we know that with probability $c$, $p \leq \mathbb{P}(\mathcal{T} \vDash \mathcal{C}_1)$. We consider that case below.

    \begin{align*}
        p &\leq \mathbb{P}(\mathcal{T} \vDash \mathcal{C}_1) \\
        &= \mathbb{P}(\tau \vDash (\mathcal{A}_1 \Rightarrow \mathcal{G}_1) \mid \tau \leftarrow \mathcal{T}) \\
        &\leq \mathbb{P}(\tau \vDash (\mathcal{A}_2 \Rightarrow \mathcal{G}_2) \mid \tau \leftarrow \mathcal{T}) && \text{(From $\mathcal{C}_1 \leq \mathcal{C}_2$)} \\
        &= \mathbb{P}(\mathcal{T} \vDash \mathcal{C}_2) \\
    \end{align*}

    Thus $\mathcal{V}_1(\mathcal{T}, c) \vdash \mathcal{C}_2$.
\end{proof}

\noindent \textbf{Theorem~\ref{def:compose_conjunction_strongmerge_vp}.1 Proof:}
\begin{proof}
    By the implicit assumption that $\mathcal{V}_1$ and $\mathcal{V}_2$ are sound validation procedures, we have $\mathbb{P}(p_1 \leq \mathbb{P}(\mathcal{T} \vDash \mathcal{C}_1)) \geq c_1$ and $\mathbb{P}(p_1 \leq \mathbb{P}(\mathcal{T} \vDash \mathcal{C}_2)) \geq c_2$. As these two events are assumed to be independent, the case where they are both true occurs with probability at least $c_1 c_2$.
    Then:
    \begin{align*}
        \mathbb{P}(\mathcal{T} \vDash \mathcal{C}_1 \parallel \mathcal{C}_2)  &= \mathbb{P}((\mathcal{T} \vDash \mathcal{C}_1) \land (\mathcal{T} \vDash \mathcal{C}_2)) &&\text{(Lemma~\ref{lemma:composition_prob})}\\
        &= 1 - \mathbb{P}(\overline{(\mathcal{T} \vDash \mathcal{C}_1)} \lor \overline{(\mathcal{T} \vDash \mathcal{C}_2)}) \\
        &\geq 1 - (\mathbb{P}(\overline{\mathcal{T} \vDash \mathcal{C}_1}) + \mathbb{P}(\overline{\mathcal{T} \vDash \mathcal{C}_2})) &&\text{(Boole's Inequality)} \\
        &= 1 - \mathbb{P}(\overline{\mathcal{T} \vDash \mathcal{C}_1}) - \mathbb{P}(\overline{\mathcal{T} \vDash \mathcal{C}_2}) \\
        &= 1 - (1 - \mathbb{P}(\mathcal{T} \vDash \mathcal{C}_1)) - (1 - \mathbb{P}(\mathcal{T} \vDash \mathcal{C}_2)) \\
        &= 1 - 1 + \mathbb{P}(\mathcal{T} \vDash \mathcal{C}_1) - 1 + \mathbb{P}(\mathcal{T} \vDash \mathcal{C}_2) \\
        &= \mathbb{P}(\mathcal{T} \vDash \mathcal{C}_1) + \mathbb{P}(\mathcal{T} \vDash \mathcal{C}_2) - 1 \\
        &\geq p_1 + p_2 - 1 \\
    \end{align*}
    Thus $\mathcal{V}_{\cup}(\mathcal{V}_1, \mathcal{V}_2, \parallel)(\mathcal{T}, c_1 c_2) \vdash (\mathcal{C}_1 \parallel \mathcal{C}_2)$.
\end{proof}

\noindent \textbf{Theorem~\ref{def:compose_conjunction_strongmerge_vp}.2 Proof:}
\begin{proof}
    By the implicit assumption that $\mathcal{V}_1$ and $\mathcal{V}_2$ are sound validation procedures, we have $\mathbb{P}(p_1 \leq \mathbb{P}(\mathcal{T} \vDash \mathcal{C}_1)) \geq c_1$ and $\mathbb{P}(p_1 \leq \mathbb{P}(\mathcal{T} \vDash \mathcal{C}_2)) \geq c_2$. As these two events are assumed to be independent, the case where they are both true occurs with probability at least $c_1 c_2$.
    Then:
    \begin{align*}
        \mathbb{P}(\mathcal{T} \vDash \mathcal{C}_1 \land \mathcal{C}_2)  &= \mathbb{P}((\mathcal{T} \vDash \mathcal{C}_1) \land (\mathcal{T} \vDash \mathcal{C}_2)) &&\text{(Lemma~\ref{lemma:conjunction_prob})}\\
        &= 1 - \mathbb{P}(\overline{(\mathcal{T} \vDash \mathcal{C}_1)} \lor \overline{(\mathcal{T} \vDash \mathcal{C}_2)}) \\
        &\geq 1 - (\mathbb{P}(\overline{\mathcal{T} \vDash \mathcal{C}_1}) + \mathbb{P}(\overline{\mathcal{T} \vDash \mathcal{C}_2})) &&\text{(Boole's Inequality)} \\
        &= 1 - \mathbb{P}(\overline{\mathcal{T} \vDash \mathcal{C}_1}) - \mathbb{P}(\overline{\mathcal{T} \vDash \mathcal{C}_2}) \\
        &= 1 - (1 - \mathbb{P}(\mathcal{T} \vDash \mathcal{C}_1)) - (1 - \mathbb{P}(\mathcal{T} \vDash \mathcal{C}_2)) \\
        &= 1 - 1 + \mathbb{P}(\mathcal{T} \vDash \mathcal{C}_1) - 1 + \mathbb{P}(\mathcal{T} \vDash \mathcal{C}_2) \\
        &= \mathbb{P}(\mathcal{T} \vDash \mathcal{C}_1) + \mathbb{P}(\mathcal{T} \vDash \mathcal{C}_2) - 1 \\
        &\geq p_1 + p_2 - 1 \\
    \end{align*}
    Thus $\mathcal{V}_{\cup}(\mathcal{V}_1, \mathcal{V}_2, \land)(\mathcal{T}, c_1 c_2) \vdash (\mathcal{C}_1 \land \mathcal{C}_2)$.
\end{proof}

\noindent \textbf{Theorem~\ref{def:compose_conjunction_strongmerge_vp}.3 Proof:}
\begin{proof}
    Using our Theorem~\ref{def:compose_conjunction_strongmerge_vp}.2 we have a sound verification procedure over $\mathcal{C}_1 \land \mathcal{C}_2$. We now show that $\mathcal{C}_1 \land \mathcal{C}_2 \leq \mathcal{C}_1 \bullet \mathcal{C}_2$.\\
    \textbf{Assumptions:}
    \begin{align*}
        \mathcal{A}_1 \land \mathcal{A}_2 &\Rightarrow \mathcal{A}_1 \lor \mathcal{A}_2 &&\text{(Trivial)}
    \end{align*}
    \textbf{Guarantees:}
    \begin{align*}
        ((\mathcal{A}_1 \Rightarrow \mathcal{G}_1) \land (\mathcal{A}_2 \Rightarrow \mathcal{G}_2)) &\Rightarrow ((\mathcal{A}_1 \land \mathcal{A}_2) \Rightarrow (\mathcal{G}_1 \land \mathcal{G}_2)) \\
        &\iff (\mathcal{G}_1 \land \mathcal{G}_2) \lor \lnot (\mathcal{A}_1 \land \mathcal{A}_2) \\
        &\iff (\mathcal{G}_1 \land \mathcal{G}_2) \lor \lnot \mathcal{A}_1 \lor \lnot \mathcal{A}_2
    \end{align*}
    Using the above and Theorem~\ref{thm:refinement_vp}, we can conclude that $\mathcal{V}_{\cup}(\mathcal{V}_1, \mathcal{V}_2, \bullet)(\mathcal{T}, c_1 c_2) \vdash (\mathcal{C}_1 \bullet \mathcal{C}_2)$.
\end{proof}

\noindent \textbf{Theorem~\ref{thm:weak_merge_vp} Proof:}
\begin{proof}
    We first show that $\forall \ \tau, \ (\tau \vDash (\mathcal{C}_1 \land \mathcal{A}_1) \lor \tau \vDash (\mathcal{C}_2 \land \lnot \mathcal{A}_1)) \Rightarrow (\tau \vDash \mathcal{C}_1 \bowtie \mathcal{C}_2)$. The assumptions are trivial, so we focus on the guarantees. If neither $\mathcal{A}_1$ or $\mathcal{A}_2$ is true, then the contract is vacuously satisfied. Otherwise, at least one must be true, so either $\mathcal{G}_1$ or $\mathcal{G}_2$ is true, satisfying the $\mathcal{G}_1 \lor \mathcal{G}_2$ portion of the contract.

    We again assume that $p_1$ and $p_2$ are lower bounds on the validity of their respective contracts (as done in previous proofs), which due to the independence of $\mathcal{V}_1$ and $\mathcal{V}_2$ occurs with probability at least $c_1 c_2$.
    \begin{align*}
        \mathbb{P}(\mathcal{T} \vDash \mathcal{C}_1 \bowtie \mathcal{C}_2) 
        &\geq (\mathcal{T} \vDash (\mathcal{C}_1 \land \mathcal{A}_1) \lor \mathcal{T} \vDash (\mathcal{C}_2 \land \lnot \mathcal{A}_1)) \\
        &= \mathbb{P}(\mathcal{T} \vDash \mathcal{C}_1 \land \mathcal{A}_1) + \mathbb{P}(\mathcal{T} \vDash \mathcal{C}_2 \land \lnot \mathcal{A}_1) && \text{(Disjoint Events)}\\
        &= \mathbb{P}(\mathcal{T} \vDash \mathcal{C}_1 \mid \mathcal{T} \vDash \mathcal{A}_1) \mathbb{P}(\mathcal{T} \vDash \mathcal{A}_1) \\
        & \quad + \mathbb{P}(\mathcal{T} \vDash \mathcal{C}_2 \mid \mathcal{T} \vDash \lnot \mathcal{A}_1) \mathbb{P}(\mathcal{T} \vDash \lnot \mathcal{A}_1) \\
        &= \mathbb{P}((\mathcal{T} \mid \mathcal{A}_1) \vDash \mathcal{C}_1) \mathbb{P}(\mathcal{T} \vDash \mathcal{A}_1) \\
        & \quad + \mathbb{P}((\mathcal{T} \mid \lnot \mathcal{A}_1) \vDash \mathcal{C}_2) (1 - \mathbb{P}(\mathcal{T} \vDash \mathcal{A}_1)) \\
        &\geq p_1 (\mathbb{P}(\mathcal{T} \vDash \mathcal{A}_1)) + p_2 (1 - \mathbb{P}(\mathcal{T} \vDash \mathcal{A}_1)) \\
    \end{align*}
\end{proof}

\noindent \textbf{Theorem~\ref{thm:weak_merge_testing_vp} Proof:}
\begin{proof}
    The proof for this verification procedure is similar to the proof for $\mathcal{V}_T$. First, we ensure that $r$ is a valid proof, returning 0 if that is not the case. By assumption $k$ is the number of traces in $\hat{\mathcal{T}_T}$ that satisfy $\mathcal{C}_2$. Furthermore, $\hat{\mathcal{T}}_P$ is composed entirely of traces that satisfy $\mathcal{C}_1$, as the set consists entirely of traces that satisfy $\mathcal{A}_1$, and thus $\mathcal{C}_1$ as we have checked that it holds using $\mathcal{V}_P$. Therefore we know that in the sample $\hat{\mathcal{T}}$ has at least $k + |\hat{\mathcal{T}_P}|$ traces that satisfy $\mathcal{C}_1$ or $\mathcal{C}_2$, and thus $\mathcal{C}_1 \bowtie \mathcal{C}_2$. Thus $\mathcal{K}_{\bowtie T}$ is a sound contract checking procedure.
\end{proof}

\section{Example Assurance Cases:}
\label{appendix:example_assurance_cases}
In this appendix we include assurance cases generated by \SPC{}, with some added line breaks (indicated by \textbackslash) and minor abbreviations for space (indicated by $\hdots$).

~\\

\noindent \textbf{Na\"ive:}
{
\begin{verbatim}
Probabilistic Contract Result:
  Component: Car(...)
  Minimum 91.55% Correctness with 99.80% Confidence
  Assumptions:
    always (self._lane is not None)
    always (((0) <= (self.speed)) and ((self.speed) <= (5.4)))
    always (((0) <= (lead_car.speed)) and ((lead_car.speed) <= (5.4)))
    always (((-(0.9)) <= ((next (self.speed)) - (self.speed)))\
        and (((next (self.speed)) - (self.speed)) <= (0.5)))
    always (((-(0.9)) <= ((next (lead_car.speed)) - (lead_car.speed)))\
        and (((next (lead_car.speed)) - (lead_car.speed)) <= (0.5)))
    ((lead_dist) > (buffer_dist)) and ((self.speed) == (0))
    always ((next (lead_dist)) == ((lead_dist) - (true_relative_speed)))
  Guarantees:
    always ((lead_dist) > (5))
  Evidence:
    Refinement Method: LeanProof: (...)
    Probabilistic Contract Result:
      Component: PerceptionSystem()
      Minimum 92.55% Correctness with 99.90% Confidence
      Assumptions:
        None
      Guarantees:
        (100.00%) always ((behind_car)\
            implies ((((lead_dist) - (0.1)) <= (dist))\
            and ((dist) <= ((lead_dist) + (0.1)))))
      Evidence:
        Simulation-Based Testing
        Sampled from Scenario 'highway.scenic (Hash=2255290405)
        3374 Verified,  559 Rejected,  0 A-Violated,  220 G-Violated
        4153 Samples, 28804.75 Seconds
        Mean Correctness: 93.88%
        Confidence Gap: 0.0266
    Contract Result:
      Component: Speedometer()
      Assumptions:
        None
      Guarantees:
        always ((speed) == (self.speed))
      Evidence:
        Assumed
    Contract Result:
      Component: ControlSystem(...)
      Assumptions:
        None
      Guarantees:
        always (((next (dist)) <= ((p_buffer_dist) + (0.1)))\
            implies ((next (throttle)) == (-(1))))
      Evidence:
        Contract Result:
          Component: ThrottleSafetyFilter(...)
          Assumptions:
            None
          Guarantees:
            always (((next (dist)) <= ((p_buffer_dist) + (0.1))) implies\
                ((next (modulated_throttle)) == (-(1))))
          Evidence:
            LeanProof: (...)
    Probabilistic Contract Result:
      Component: CarActionControls()
      Minimum 99.00% Correctness with 99.90% Confidence
      Assumptions:
        None
      Guarantees:
        always (((throttle) == (-(1))) implies\
            (((next (self.speed)) == (0)) or\
            ((next (self.speed)) == ((self.speed) - (0.9)))))
      Evidence:
        Assumed
\end{verbatim}}

\noindent\textbf{Optimized:}
{
\begin{verbatim}
Probabilistic Contract Result:
  Component: Car(...)
  Minimum 94.59% Correctness with 99.80% Confidence
  Assumptions:
    always (self._lane is not None)
    always (((0) <= (self.speed)) and ((self.speed) <= (5.4)))
    always (((0) <= (lead_car.speed)) and ((lead_car.speed) <= (5.4)))
    always (((-(0.9)) <= ((next (self.speed)) - (self.speed)))\
        and (((next (self.speed)) - (self.speed)) <= (0.5)))
    always (((-(0.9)) <= ((next (lead_car.speed)) - (lead_car.speed)))\
        and (((next (lead_car.speed)) - (lead_car.speed)) <= (0.5)))
    ((lead_dist) > (buffer_dist)) and ((self.speed) == (0))
    always ((next (lead_dist)) == ((lead_dist) - (true_relative_speed)))
  Guarantees:
    always ((lead_dist) > (5))
  Evidence:
    Refinement Method: LeanProof: (...)
    Probabilistic Contract Result:
      Component: PerceptionSystem()
      Minimum 95.59% Correctness with 99.90% Confidence
      Assumptions:
        None
      Guarantees:
        always ((behind_car) implies\
            ((((lead_dist) - (0.1)) <= (dist))\
            and ((dist) <= ((lead_dist) + (0.1)))))
      Evidence:
        Refinement Method: LeanProof: (...)
        Conjunction Result:
        Simulation-Based Testing
            Sampled from Scenario 'NONE (Hash=NONE)
            5322 Verified,  810 Rejected,  0 A-Violated,  197 G-Violated
            6329 Samples, 28803.76 Seconds
            Mean Correctness: 96.43%
            Confidence Gap: 0.0166
        Sub Result (Correctness=1.00):
        Contract Result:
          Component: PerceptionSystem()
          Assumptions:
            (((params['weather']) == (0))\
                or ((params['weather']) == (1)))\
                and ((params['lead_car_width']) >= (1.8))
          Guarantees:
            always ((behind_car) implies\
                ((((lead_dist) - (0.1)) <= (dist))\
                and ((dist) <= ((lead_dist) + (0.1)))))
          Evidence:
            Refinement Method: LeanProof: (...)
            Contract Result:
              Component: RadarDistanceSystem()
              Assumptions:
                (params['lead_car_width']) >= (1.8)
              Guarantees:
                always ((behind_car) implies\
                    ((((lead_dist) - (0.1)) <= (dist))\
                    and ((dist) <= ((lead_dist) + (0.1)))))
              Evidence:
                Assumed
            Contract Result:
              Component: LaserDistanceSystem()
              Assumptions:
                ((params['weather']) == (0))\
                    or ((params['weather']) == (1))
              Guarantees:
                always ((behind_car) implies\
                    ((((lead_dist) - (0.1)) <= (dist))\
                    and ((dist) <= ((lead_dist) + (0.1)))))
              Evidence:
                Assumed
            Contract Result:
              Component: MedianDistanceFilter()
              Assumptions:
                None
              Guarantees:
                always ((out_dist) == \
                    (min((min((max((dist1), (dist2))),\
                    (max((dist1), (dist2))))),\
                    (max((dist2), (dist3))))))
              Evidence:
                LeanProof: (...)
        Sub Result (Correctness=0.93):
        Probabilistic Contract Result:
          Component: PerceptionSystem()
          Minimum 93.34% Correctness with 99.90% Confidence
          Assumptions:
            (  0.00%) not ((((params['weather']) == (0))\
                or ((params['weather']) == (1)))\
                and ((params['lead_car_width']) >= (1.8)))
          Guarantees:
            (100.00%) always ((behind_car) implies\
                ((((lead_dist) - (0.1)) <= (dist))\
                and ((dist) <= ((lead_dist) + (0.1)))))
          Evidence:
            Simulation-Based Testing
            Sampled from Scenario 'highway.scenic (Hash=2255290405)
            3446 Verified,  534 Rejected,  0 A-Violated,  197 G-Violated
            4177 Samples, 28803.76 Seconds
            Mean Correctness: 94.59%
            Confidence Gap: 0.0250
        
    Contract Result:
      Component: Speedometer()
      Assumptions:
        None
      Guarantees:
        always ((speed) == (self.speed))
      Evidence:
        Assumed
    Contract Result:
      Component: ControlSystem(...)
      Assumptions:
        None
      Guarantees:
        always (((next (dist)) <= ((p_buffer_dist) + (0.1)))\
            implies ((next (throttle)) == (-(1))))
      Evidence:
        Contract Result:
          Component: ThrottleSafetyFilter(...)
          Assumptions:
            None
          Guarantees:
            always (((next (dist)) <= ((p_buffer_dist) + (0.1)))\
                implies ((next (modulated_throttle)) == (-(1))))
          Evidence:
            LeanProof: (...)
    Probabilistic Contract Result:
      Component: CarActionControls()
      Minimum 99.00% Correctness with 99.90% Confidence
      Assumptions:
        None
      Guarantees:
        always (((throttle) == (-(1))) implies\
            (((next (self.speed)) == (0))\
            or ((next (self.speed)) == ((self.speed) - (0.9)))))
      Evidence:
        Assumed
\end{verbatim}}

\noindent \textbf{Na\"ive:}
{
\begin{verbatim}
Probabilistic Contract Result:
  Component: InspectionDrone(wall_y=10.0, floor_alt=4.0, wind_limit=6.0, alpha=1.0, d_min=1.0, margin_buffer=2.2, approach_speed=2.0, descent_speed=1.5, cruise_speed=3.0)
  Minimum 99.61% Correctness with 99.90% Confidence
  Assumptions:
    (  0.00%) ((true_wall) >= (5.5)) and ((true_floor) >= (5.0))
  Guarantees:
    (  0.00%) always ((min((true_wall), (true_floor))) > (1.0))
    (  0.00%) always ((not (next (False))) implies (((true_wall) >= (5.5)) implies ((wall_movement) >= ((0.099002112) * ((2.0) - (1.4))))))
    (  0.00%) always ((not (next (False))) implies (((true_floor) >= (5.0)) implies ((floor_movement) >= ((0.099002112) * ((1.5) - (0.8))))))
  Evidence:
    Simulation-Based Testing
    Sampled from Scenario 'corridor.scenic (Hash=1143642744)
    1752 Verified,  0 Rejected,  0 A-Violated,  0 G-Violated
    1752 Samples, 28809.96 Seconds
    Mean Correctness: 100.00%
    Confidence Gap: 0.0043
    Individual Guarantee Evidence:
      corridor clearance: 1752/1752 passed, 0 failed, correctness >= 99.61% at 99.90% confidence
      wall progress: 1752/1752 passed, 0 failed, correctness >= 99.61% at 99.90% confidence
      floor progress: 1752/1752 passed, 0 failed, correctness >= 99.61% at 99.90% confidence
\end{verbatim}
}

\noindent \textbf{Compositional:}
{
\begin{verbatim}
Probabilistic Contract Result:
  Component: InspectionDrone(wall_y=10.0, floor_alt=4.0, wind_limit=6.0, alpha=1.0, d_min=1.0, margin_buffer=2.2, approach_speed=2.0, descent_speed=1.5, cruise_speed=3.0)
  Minimum 99.81% Correctness with 99.90% Confidence
  Assumptions:
    ((true_wall) >= (5.5)) and ((true_floor) >= (5.0))
  Guarantees:
    always ((min((true_wall), (true_floor))) > (1.0))
    always ((not (next (False))) implies (((true_wall) >= (5.5)) implies ((wall_movement) >= ((0.099002112) * ((2.0) - (1.4))))))
    always ((not (next (False))) implies (((true_floor) >= (5.0)) implies ((floor_movement) >= ((0.099002112) * ((1.5) - (0.8))))))
  Evidence:
    Refinement Method: LeanProof: ('LeanLTL/Examples/Scenic/CorridorActuation/CorridorContainmentRefinement')
    Conjunction Result:
    3718 Verified,  0 Rejected,  0 A-Violated,  0 G-Violated
        3718 Samples, 28807.34 Seconds
        Mean Correctness: 100.00%
        Confidence Gap: 0.0020
        Individual Guarantee Evidence:
          guarantee 1: 3718/3718 passed, 0 failed, correctness >= 99.81% at 99.90% confidence
          guarantee 2: 3718/3718 passed, 0 failed, correctness >= 99.81% at 99.90% confidence
    Sub Result (Correctness=1.00):
    Contract Result:
      Component: InspectionDrone(wall_y=10.0, floor_alt=4.0, wind_limit=6.0, alpha=1.0, d_min=1.0, margin_buffer=2.2, approach_speed=2.0, descent_speed=1.5, cruise_speed=3.0)
      Assumptions:
        ((params['gnssQuality']) == (0)) and ((((params['windAlong']) * (params['windAlong'])) + ((params['windCross']) * (params['windCross']))) <= (36.0)) and (((params['windCross']) * (params['windCross'])) <= (25.0)) and (((params['windVert']) * (params['windVert'])) <= (4.0))
      Guarantees:
        (((true_wall) >= (5.5)) and ((true_floor) >= (5.0))) implies (always ((min((true_wall), (true_floor))) > (1.0)))
        (((true_wall) >= (5.5)) and ((true_floor) >= (5.0))) implies (always ((not (next (False))) implies (((true_wall) >= (5.5)) implies ((wall_movement) >= ((0.099002112) * ((2.0) - (1.4)))))))
        (((true_wall) >= (5.5)) and ((true_floor) >= (5.0))) implies (always ((not (next (False))) implies (((true_floor) >= (5.0)) implies ((floor_movement) >= ((0.099002112) * ((1.5) - (0.8)))))))
      Evidence:
        Refinement Method: LeanProof: ('LeanLTL/Examples/Scenic/CorridorActuation/CorridorContainmentKnownRefinement')
        Contract Result:
          Component: CorridorPerception(wall_y=10.0, floor_alt=4.0, wind_limit=6.0)
          Assumptions:
            ((params['gnssQuality']) == (0)) and ((((params['windAlong']) * (params['windAlong'])) + ((params['windCross']) * (params['windCross']))) <= (36.0))
          Guarantees:
            always ((((true_wall) - (0.3)) <= (wall_margin)) and ((wall_margin) <= ((true_wall) + (0.3))))
            always ((((true_floor) - (0.3)) <= (floor_margin)) and ((floor_margin) <= ((true_floor) + (0.3))))
          Evidence:
            Refinement Method: LeanProof: ('LeanLTL/Examples/Scenic/CorridorActuation/AccurateMarginsKnownRefinement')
            Contract Result:
              Component: GnssPositionSystem()
              Assumptions:
                (params['gnssQuality']) == (0)
              Guarantees:
                always ((((self.trueLateral) - (0.3)) <= (gnss_lateral)) and ((gnss_lateral) <= ((self.trueLateral) + (0.3))))
                always ((((self.trueAltitude) - (0.3)) <= (gnss_altitude)) and ((gnss_altitude) <= ((self.trueAltitude) + (0.3))))
              Evidence:
                Assumed
            Contract Result:
              Component: VioPositionSystem(wind_limit=6.0)
              Assumptions:
                (((params['windAlong']) * (params['windAlong'])) + ((params['windCross']) * (params['windCross']))) <= (36.0)
              Guarantees:
                always ((((self.trueLateral) - (0.3)) <= (vio_lateral)) and ((vio_lateral) <= ((self.trueLateral) + (0.3))))
                always ((((self.trueAltitude) - (0.3)) <= (vio_altitude)) and ((vio_altitude) <= ((self.trueAltitude) + (0.3))))
              Evidence:
                Assumed
            Contract Result:
              Component: CorridorMarginEstimator(wall_y=10.0, floor_alt=4.0)
              Assumptions:
                None
              Guarantees:
                always (((min(((10.0) - (gnss_lateral)), ((10.0) - (vio_lateral)))) <= (wall_margin)) and ((wall_margin) <= (max(((10.0) - (gnss_lateral)), ((10.0) - (vio_lateral))))))
                always (((min(((gnss_altitude) - (4.0)), ((vio_altitude) - (4.0)))) <= (floor_margin)) and ((floor_margin) <= (max(((gnss_altitude) - (4.0)), ((vio_altitude) - (4.0))))))
              Evidence:
                LeanProof: ('LeanLTL/Examples/Scenic/CorridorActuation/CorridorMarginSemantics')
        Contract Result:
          Component: ChannelGovernor(alpha=1.0, d_min=1.0, margin_buffer=2.2, max_rate=2.0)
          Assumptions:
            None
          Guarantees:
            always ((cmd) <= ((1.0) * (((margin) - (1.0)) - (2.2))))
            always ((cmd) <= (2.0))
            always (((margin) >= (((1.0) + (2.2)) + ((2.0) / (1.0)))) implies ((cmd) == (2.0)))
            always ((prev_cmd) <= (2.0))
            always ((next (prev_cmd)) == (cmd))
          Evidence:
            LeanProof: ('LeanLTL/Examples/Scenic/CorridorActuation/WallGovernorSafety')
        Contract Result:
          Component: ChannelGovernor(alpha=1.0, d_min=1.0, margin_buffer=2.2, max_rate=1.5)
          Assumptions:
            None
          Guarantees:
            always ((cmd) <= ((1.0) * (((margin) - (1.0)) - (2.2))))
            always ((cmd) <= (1.5))
            always (((margin) >= (((1.0) + (2.2)) + ((1.5) / (1.0)))) implies ((cmd) == (1.5)))
            always ((prev_cmd) <= (1.5))
            always ((next (prev_cmd)) == (cmd))
          Evidence:
            LeanProof: ('LeanLTL/Examples/Scenic/CorridorActuation/FloorGovernorSafety')
        Contract Result:
          Component: DroneActuation(cruise_speed=3.0)
          Assumptions:
            (((params['windCross']) * (params['windCross'])) <= (25.0)) and (((params['windVert']) * (params['windVert'])) <= (4.0))
          Guarantees:
            always ((not (next (False))) implies ((((0.099002112) * ((cmd_closing) - (1.4))) <= (wall_movement)) and ((wall_movement) <= ((0.099002112) * ((max((cmd_closing), (prev_closing))) + (1.8))))))
            always ((not (next (False))) implies ((((0.099002112) * ((cmd_descent) - (0.8))) <= (floor_movement)) and ((floor_movement) <= ((0.099002112) * ((max((cmd_descent), (prev_descent))) + (1.0))))))
          Evidence:
            Assumed
    Sub Result (Correctness=1.00):
    Probabilistic Contract Result:
      Component: InspectionDrone(wall_y=10.0, floor_alt=4.0, wind_limit=6.0, alpha=1.0, d_min=1.0, margin_buffer=2.2, approach_speed=2.0, descent_speed=1.5, cruise_speed=3.0)
      Minimum 99.60% Correctness with 99.90% Confidence
      Assumptions:
        (  0.00%) not (((params['gnssQuality']) == (0)) and ((((params['windAlong']) * (params['windAlong'])) + ((params['windCross']) * (params['windCross']))) <= (36.0)) and (((params['windCross']) * (params['windCross'])) <= (25.0)) and (((params['windVert']) * (params['windVert'])) <= (4.0)))
      Guarantees:
        (  0.00%) (((true_wall) >= (5.5)) and ((true_floor) >= (5.0))) implies (always ((min((true_wall), (true_floor))) > (1.0)))
        (  0.00%) (((true_wall) >= (5.5)) and ((true_floor) >= (5.0))) implies (always ((not (next (False))) implies (((true_wall) >= (5.5)) implies ((wall_movement) >= ((0.099002112) * ((2.0) - (1.4)))))))
        (  0.00%) (((true_wall) >= (5.5)) and ((true_floor) >= (5.0))) implies (always ((not (next (False))) implies (((true_floor) >= (5.0)) implies ((floor_movement) >= ((0.099002112) * ((1.5) - (0.8)))))))
      Evidence:
        Simulation-Based Testing
        Sampled from Scenario 'corridor.scenic (Hash=1143642744)
        1738 Verified,  0 Rejected,  0 A-Violated,  0 G-Violated
        1738 Samples, 28807.34 Seconds
        Mean Correctness: 100.00%
        Confidence Gap: 0.0044
        Individual Guarantee Evidence:
          corridor clearance: 1738/1738 passed, 0 failed, correctness >= 99.60% at 99.90% confidence
          wall progress: 1738/1738 passed, 0 failed, correctness >= 99.60% at 99.90% confidence
          floor progress: 1738/1738 passed, 0 failed, correctness >= 99.60% at 99.90% confidence
\end{verbatim}
}

\section{Additional Case Study Figures}
We include additional experimental results to aid in artifact evaluation showing the correctness probability as a factor of the number of simulations run, as this is more reproducible on different systems than a time budget.

\begin{figure}[H]
    \centering
    \includegraphics[width=0.8\textwidth]{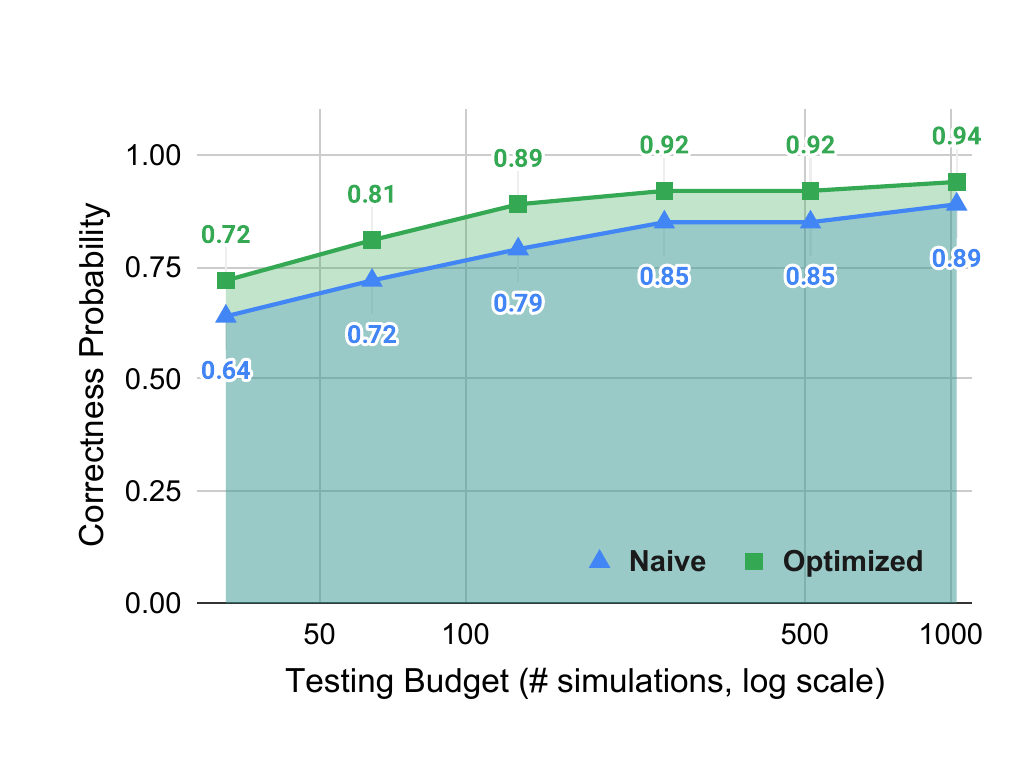}
    \caption{Probability lower bounds on the overall system contract (\textsc{Keeps Distance}) obtained with \SPC{}, as a function of number of simulations run.}
    \label{fig:contract_probability_samples}
\end{figure}

\begin{figure}[H]
    \centering
    \includegraphics[width=0.8\textwidth]{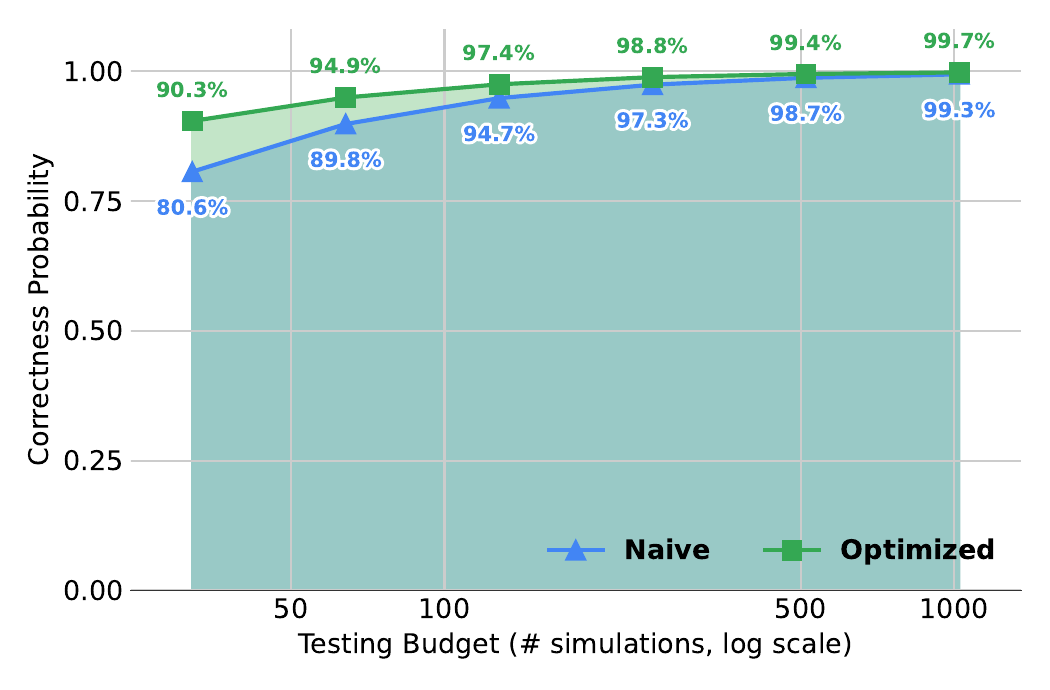}
    \caption{Probability lower bounds on the overall system contract (\textsc{CorridorContainment}) obtained with \SPC{}, as a function of number of simulations run.}
    \label{fig:drone_probability_samples}
\end{figure}

\begin{figure}[tb]
    \centering
    \includegraphics[width=\linewidth]{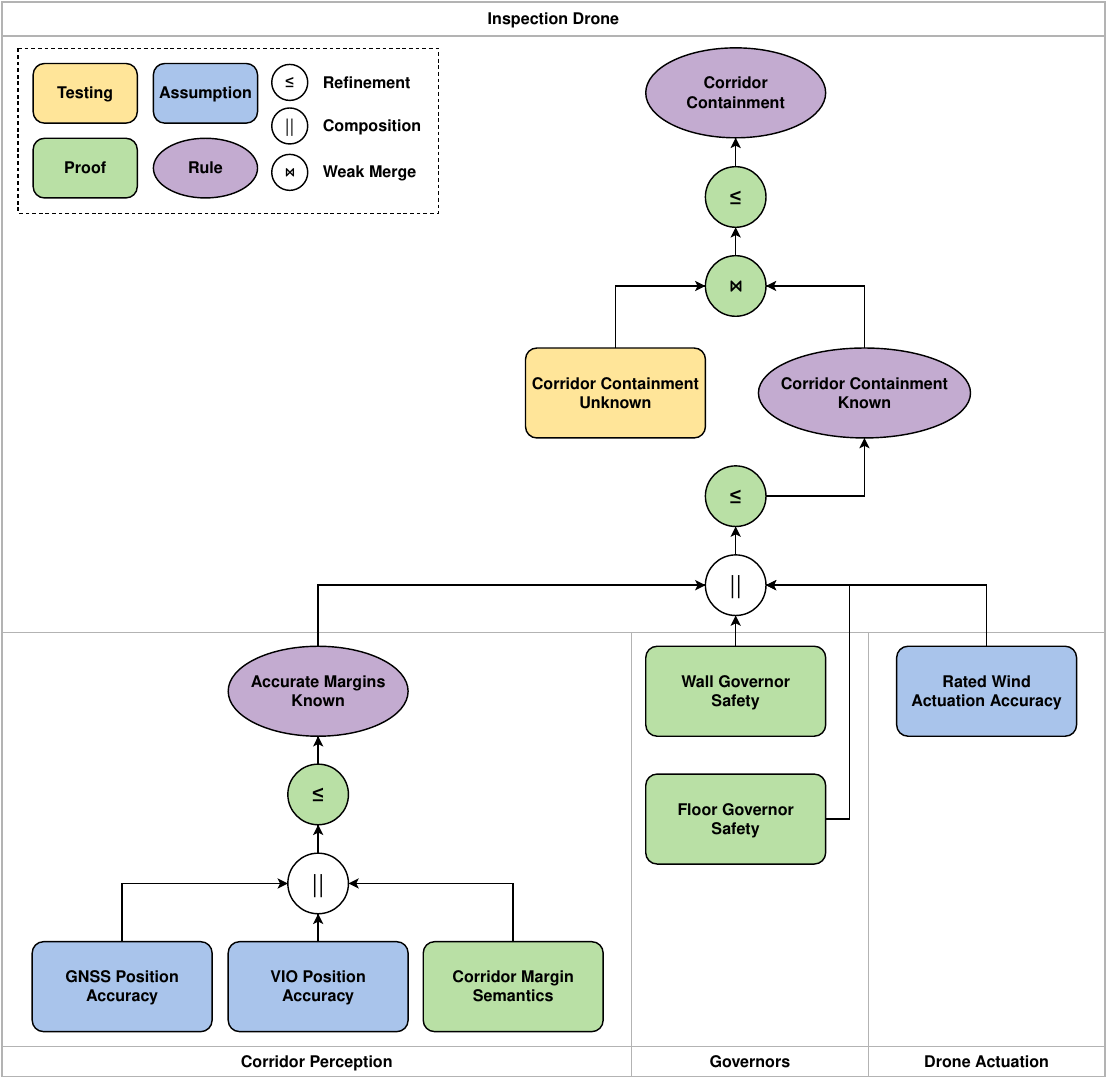}
    \caption{The structure of the drone assurance case.}
    \label{fig:drone_assurance_case}
\end{figure}

\subsection{Drone Case Study Parameters}
\label{appendix:drone_parameters}

\begin{table}[h]
\centering
\begin{tabular}{lc}
\hline
Parameter & Value \\
\hline
Minimum clearance from wall and floor (safety) & $1\,\mathrm{m}$ \\
Controller target clearance & $3.2\,\mathrm{m}$ \\
Maximum commanded approach rate (wall / floor) & $2$ / $1.5\,\mathrm{m/s}$ \\
Maximum GNSS error under poor reception & $1.5\,\mathrm{m}$ \\
VIO rated horizontal wind & $\leq 6\,\mathrm{m/s}$ \\
Sensor error bound under rated conditions & $0.3\,\mathrm{m}$ \\
Actuation rated crosswind / vertical wind & $\leq 5$ / $\leq 2\,\mathrm{m/s}$ \\
\hline
\end{tabular}
\caption{Parameters for the drone case study.}
\end{table}

\end{document}